\documentclass[reprint, superscriptaddress, amsmath, amssymb, aps, nofootinbib, floatfix]{revtex4-2}
\usepackage[utf8]{inputenc}
\usepackage{amsmath}
\usepackage{amsthm}
\usepackage[normalem]{ulem}
\usepackage{graphicx}
\usepackage{soul}
\usepackage{appendix}
\usepackage{ulem}
\usepackage{xcolor}
\usepackage[T1]{fontenc}
\usepackage{braket}
\usepackage{mathtools}
\graphicspath{{}}

\begin{document}
% \title{Two-Body Problem with Contact Interaction in a Harmonic Trap and Optical Lattice}
\title{Two-Body Kapitza--Dirac Scattering of One-Dimensional Ultracold Atoms}
\date{\today}
\author{A. Becker}
\email{andre.becker@uni-hamburg.de}
\affiliation{Center for Optical Quantum Technologies, Department of Physics, University of Hamburg, 
Luruper Chaussee 149, 22761 Hamburg Germany}
\affiliation{The Hamburg Centre for Ultrafast Imaging,
University of Hamburg, Luruper Chaussee 149, 22761 Hamburg, Germany}
\author{G. M. Koutentakis}
\email{georgios.koutentakis@ist.ac.at}
\affiliation{Institute of Science and Technology Austria (ISTA), am Campus 1,\\ 3400 Klosterneuburg, Austria}
\author{P. Schmelcher}
\email{peter.schmelcher@uni-hamburg.de}
\affiliation{Center for Optical Quantum Technologies, Department of Physics, University of Hamburg, 
Luruper Chaussee 149, 22761 Hamburg Germany}
\affiliation{The Hamburg Centre for Ultrafast Imaging,
University of Hamburg, Luruper Chaussee 149, 22761 Hamburg, Germany}
\begin{abstract}
Kapitza--Dirac scattering, the diffraction of matter waves from a standing light field, is widely utilized in ultracold gases, but its behavior in the strongly interacting regime is an open question. Here we develop a numerically-exact two-body description of Kapitza--Dirac scattering for two contact-interacting atoms in a one-dimensional harmonic trap subjected to a pulsed optical lattice, enabling us to obtain the numerically exact dynamics. We map how interaction strength, lattice depth, lattice wavenumber, and pulse duration reshape the diffraction pattern, leading to an interaction-dependent population redistribution in real and momentum-space. By comparing the exact dynamics to an impulsive sudden-approximation description, we delineate the parameter regimes where it remains accurate and those, notably at strong attraction and small lattice wavenumber, where it fails. Our results provide a controlled few-body benchmark for interacting Kapitza--Dirac scattering and quantitative guidance for Kapitza--Dirac-based probes of ultracold atomic systems.

\end{abstract}

% HAMILTONIAN
\maketitle
\section{Introduction}

Originally proposed in the context of electron diffraction by standing light \cite{KapitzaDirac1933}, the Kapitza--Dirac effect constitutes one of the earliest theoretical proposals for the explicit demonstration of the wave nature of matter \cite{Batelaan2000}. It was first realized in atomic beams once intense, coherent laser sources became available \cite{Gould1986, Martin1988, Ovchinnikov1999}, while the original proposal was achieved experimentally only recently \cite{Freimund2001, Freimund2001B}. Depending on pulse duration and field strength, the Kapitza--Dirac scattering interpolates between the Raman–Nath regime \cite{RamanNath1935}, where short and strong pulses populate many momentum orders, and the Bragg regime \cite{Bragg1913}, where long and weak pulses couple only a few resonant states under strict energy and momentum conservation. This crossover can be described in terms of atoms moving in a pulsed optical lattice, using Bloch-state approaches \cite{Champenois2001} or effective few-level models \cite{Mueller2008}.

With the advent of laser cooling and trapping, Kapitza--Dirac scattering has become a versatile tool for manipulating and probing ultracold atomic gases \cite{Cronin2009, Betalaan2007, SANCHO2015}. Diffraction of Bose–Einstein condensates has been used to characterize lattice depths with high precision \cite{Gadway2009}, measure polarizabilities and determine the tune-out wavelength \cite{Ratkata2021}, while Bragg-type protocols have provided sensitive probes of coherence and correlations in optical lattices \cite{Gadway2012}. More recently, Kapitza--Dirac scattering has been employed in strongly interacting Fermi gases to monitor the response of molecular condensates across Fano–Feshbach resonances \cite{Liang2022} and through the BEC–BCS crossover \cite{Jin2024}. Momentum-space lattice implementations further broadened the conceptual framework for understanding higher-order diffraction as effective tunneling dynamics \cite{gadway2015,Meyer2016}. In these settings, the interplay of interactions, confinement, and lattice-induced diffraction is central for interpreting experimental data.

Indeed, mean-field and many-body effects are known to modify Bragg spectra and momentum-resolved response \cite{Stenger1999,Blakie2002}, and nonadiabatic dynamics can induce deviations from simple Rabi-type population transfer \cite{Reeves2015}.
On the theoretical side, mean-field descriptions, effective few-mode models, and many-body simulations have all been used to incorporate interactions into Kapitza--Dirac–type dynamics \cite{Fitzek2020}. However, these approaches are either approximate by construction—relying on mean-field factorization, truncated mode expansions, or perturbative treatments—or so numerically demanding that they obscure the direct connection between microscopic parameters and observable diffraction patterns. For strongly interacting systems in tight traps, an exact few-body description is therefore essential: it provides a controlled benchmark for assessing the accuracy of such approximations and a transparent framework for disentangling the roles of interactions, confinement, and lattice parameters. The minimal system of two strongly correlated confined atoms, analytically solvable by the methods of Busch {\it et al.} \cite{Busch1998} and later developments \cite{Budewig2019, Bougas2019}, offers precisely this kind of foundation. Together with the deterministic preparation of few-atom systems \cite{Serwane2011} and the observation of tunable two-body bound states in one-dimensional geometries \cite{Zuern2012}, this highlights the need for a rigorous two-body treatment of Kapitza--Dirac scattering that goes beyond existing approximate or purely numerical many-body approaches and can serve as a rigorous quantitative benchmark for them.

In this work, we address this gap in the literature by developing a detailed theoretical description of Kapitza--Dirac scattering in an interacting two-body system confined in a one-dimensional harmonic trap. We consider two atoms in their exchange symmetric state (bosons regardless of internal state and fermions in a hyperfine singlet state) with tunable contact interactions and subject them to a pulsed optical lattice. The exact Busch solution provides the eigenstates and spectrum of the trapped two-body problem, which we use as a basis to incorporate the lattice potential, including the coupling between center-of-mass and relative motion. We then solve the time-dependent Schrödinger equation by full configuration interaction, allowing us to follow the dynamics during the pulse. This framework yields a numerically exact benchmark for interacting Kapitza--Dirac scattering in which we can systematically chart how interaction strength, lattice depth, lattice momentum, and pulse duration shape the ensuing diffraction pattern.

To connect directly to experimentally accessible observables, we compute the spatial one- and two-body densities and the momentum distributions, directly accessible via in situ and time-of-flight measurements in state-of-the-art experiments \cite{BergschneiderThesis2017, BergschneiderKlinkhamer2018, BergschneiderKlinkhamer2019, PreissBecher2019}. By comparing the exact dynamics with an impulsive description based on the sudden approximation and evaluating the corresponding fidelity, we identify parameter regimes where the sudden approximation accurately captures Kapitza--Dirac scattering and regimes, particularly at strong attraction and for small lattice wave numbers, where it fails. In this way, our minimal two-body model both clarifies the role of interactions and confinement in Kapitza--Dirac diffraction and provides a quantitative benchmark for interpreting and designing Kapitza--Dirac-based experiments on few- and many-body systems.

The structure of this work is as follows. Section~\ref{sec:theory} introduces the theoretical framework, discusses the two-body basis, and outlines the employed calculation methodology and observables. 
%We then derive the one-body and two-body densities and their associated momentum distributions, which serve as key observables for characterizing the scattering process.
Section~\ref{sec:Results} presents our main results on the manifestation of Kapitza--Dirac diffraction and the dependence of the dynamics on system parameters and compares them to the sudden approximation. We summarize our findings in Sect.~\ref{sec:Summary_Outlook} and provide an outlook. Additional technical details, including convergence properties (Appendix~\ref{sec:conv_analysis}), analytical expressions for displacement-operator matrix elements (Appendix~\ref{sec_app:Mat_Elemenets_and_Displacement_Operator}), and the transformation between center-of-mass and laboratory frames (Appendix~\ref{sec_app:Rotation_to_Lab_frame}), are provided in the Appendices.
\section{Theoretical Framework}
\label{sec:theory}
\begin{figure}
    \centering
    \includegraphics[width=1.\linewidth]{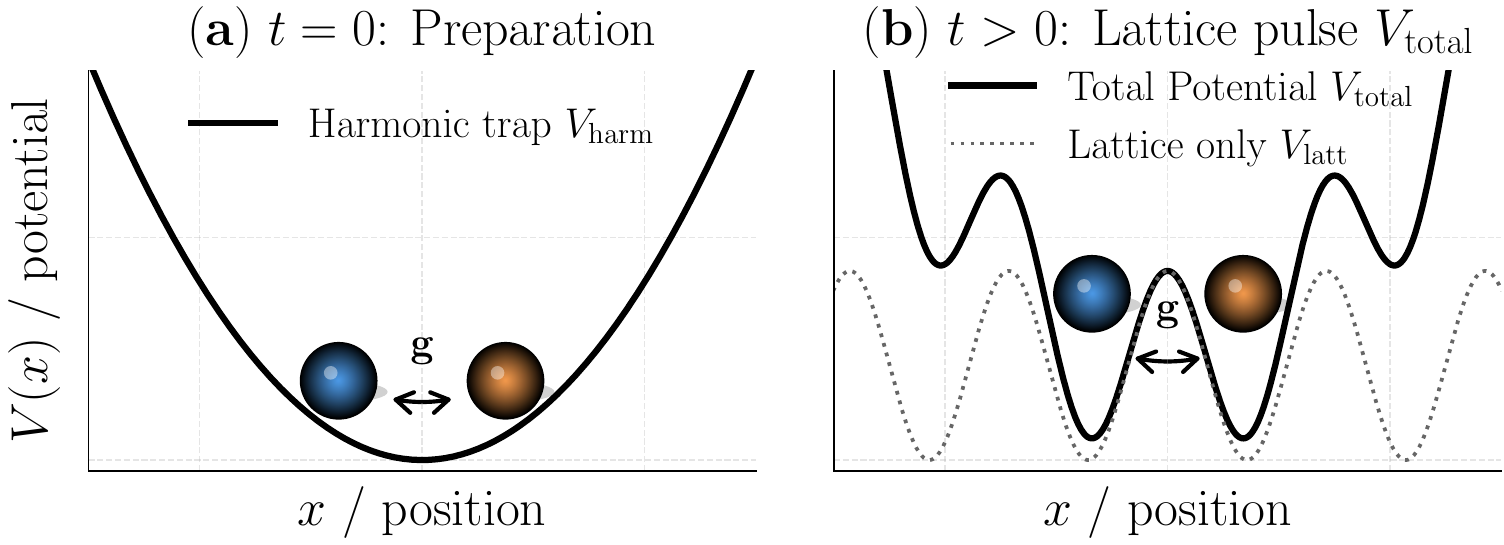}
    \caption{(a) Two atoms (blue and orange spheres) in a harmonic trap at $t=0$, interacting via a contact interaction of strength $g$ at the trap center.
    (b) For $t>0$, an additional optical lattice potential is applied on top of the harmonic confinement. The total potential (solid) and lattice potential (dashed) are shown separately for clarity.}
    \label{fig:setup}
\end{figure}
Understanding Kapitza--Dirac diffraction in interacting ultracold systems poses substantial theoretical and computational challenges. In order to reduce the complexity and computational cost due to many-body effects, we consider the simplest non-trivial system. This consists of two interacting atoms confined in one dimension by a parabolic potential as depicted in Fig.~\ref{fig:setup}. This minimal model offers two important advantages: First, we can purely focus on interaction effects and coherent momentum transfer from the light field. Second, it allows a fully quantum-mechanical, time-dependent treatment of the Kapitza--Dirac process without simplifying assumptions like mean-field or semi-classical approximations. Herewith we will adopt the fermionic notation of a spin-$\uparrow$ and a spin-$\downarrow$ particle with the same mass, $m = m_{\uparrow} = m_{\downarrow}$, however, as we will argue later on, our results are more broadly applicable. We express our one-dimensional Hamiltonian as
\begin{align}
H &=
\sum_{\sigma=\uparrow,\downarrow}
\left[
-\frac{\hbar^{2}}{2m}\frac{\partial^{2}}{\partial x_{\sigma}^{2}}
+\frac{1}{2}m\omega^{2}x_{\sigma}^{2}
+U_{0}\cos^{2}\!\bigl(k_{\rm lat}x_{\sigma}\bigr)
\right]
\nonumber\\
&\quad + \sqrt{2}\,g\,\delta\!\bigl(x_{\uparrow}-x_{\downarrow}\bigr).
\label{total_Hamiltonian}
\end{align}
where $x_{\sigma}$ denotes the coordinate of the $\sigma$-th particle, $k_{\rm lat} = \pi/a$ is the lattice wave vector with $a$ the lattice constant, $U_0$ the lattice depth, and $g$ the contact interaction strength between the two atoms. The contact interaction provides an effective low-energy description of a Fano–Feshbach resonance, where the scattering length can be tuned experimentally via an external magnetic field \cite{Huang1957, Feshbach1964, Bloch2008, Chin2010}. The prefactor $\sqrt{2}$ is introduced for later convenience when transforming the Hamiltonian to center-of-mass and relative coordinates. Subsequently, we express all quantities in harmonic-oscillator units, with the characteristic length scale given by $\ell = \sqrt{\hbar/(m\omega)}$.

For the unperturbed Hamilton operator, i.e. without the lattice potential $U_0=0$, the solution is well known \cite{Busch1998}. However, to apply these solutions, we need to perform a transformation to relative and center-of-mass (CM) coordinates
\begin{equation}
r = \frac{1}{\sqrt{2}}(x_{\uparrow} - x_{\downarrow}), \quad R = \frac{1}{\sqrt{2}}(x_{\uparrow} + x_{\downarrow}).
\label{eqn:coordinate_transformation}
\end{equation}
In this frame the Hamiltonian decomposes as $H = H_{\rm CM} + H_{\rm rel} + H_{\rm lat}$, with
\begin{align}
H_{\text{CM}} &= -\frac{\hbar^2}{2m}\frac{\partial^2}{\partial R^2} + \frac{1}{2}m\omega^2R^2, \\
H_{\text{rel}} &= -\frac{\hbar^2}{2m}\frac{\partial^2}{\partial r^2} + \frac{1}{2}m\omega^2r^2 + g \delta(r), \\
H_{\text{lat}} &= U_0 \left[ \cos{(\sqrt{2} k_{\rm lat} r)} \cos{(\sqrt{2} k_{\rm lat} R)} + 1 \right].
\label{eqn:lattice_potential_ham}
\end{align}
Let us now briefly reiterate the statistics discussion. In the CM frame, the statistics of the particles dictate whether the relative wavefunction is even or odd. As is obvious by the above expression both $H_{\text{rel}}$ and $H_{\text{lat}}$ conserve the parity of the relative state as they are invariant by the $r \to -r$ transformation. Therefore, our results equivalently apply for any species allowing a parity symmetric relative state --spinless/spinful bosons and fermions in a spin-singlet state-- as the parity odd states can never be accessed by parity even ones. Notice that parity odd states do not experience interaction since they possess a node at $r = 0$ and thus are not considered in the following.

For $U_0=0$ the system is strictly separable, thus the time-evolved states (and the eigenstates) factorize, $\Psi(R,r;t)=\psi^{\rm CM}(R;t)\,\psi^{\rm rel}(r;t)$.
When $U_0\neq 0$, the term $H_{\rm lat}$ couples center-of-mass and relative motion and states are no longer factorizable. Nevertheless, the direct products of $U_0 = 0$ eigenstates $\{\psi^{\rm CM}_n(R)\otimes\psi^{\rm rel}_k(r;g)\}$ provide a convenient basis to evaluate the coupling matrix. The calculation of the coupling matrix stemming from $H_{\rm lat}$ is performed in the next sections.
\subsection{Interacting harmonic oscillator in 1D: the Busch-Englert-Rza{\. z}ewski-Wilkens solution}

The two-body harmonic oscillator with a contact interaction is exactly solvable in all relevant dimensions \cite{Busch1998, Idzsiaszek2005}. Thus, Eq.~\eqref{total_Hamiltonian} is exactly solvable for $U_0=0$, and the corresponding eigenstates form the natural basis for our treatment of the lattice-induced dynamics.

In the CM frame, the CM eigenfunctions are harmonic-oscillator (HO) states $\psi_{\rm CM}(R) = \psi_n^{\rm HO}(R)$, while the relative eigenfunctions depend on the interaction strength $g$.
Odd relative modes are unaffected by the contact interaction, because
$\psi_{2n+1}^{\rm rel}(-r;g) = - \psi_{2n+1}^{\rm rel}(r;g)$ implies $\psi_{2n+1}^{\rm rel}(0;g)=0$, and one has
$\psi_{2n+1}^{\rm rel}(r;g) = \psi_{2 n + 1}^{\rm HO}(r)$.
In contrast, even relative modes are interaction-dependent. Busch {\rm et al.}\ showed that they form a family of states parametrized by an effective order parameter $\epsilon_n(g)$,
\begin{equation}
\psi_{2 n}^{\rm B}(r; g)=\frac{A_{n}(g) \Gamma(-\epsilon_{n}(g))}{2 \sqrt{\pi \ell}} U\left( -\epsilon_{n}(g), \frac{1}{2}, \frac{r^2}{\ell^2} \right) e^{-\frac{r^2}{2 \ell^2}},
\label{eq:busch_rel_func_main_text}
\end{equation}
where $U(a,b,z)$ is Tricomi’s confluent hypergeometric function \cite{gradshteyn_ryzhik_2014} and $\Gamma(z)$ is the Gamma function.
The interaction enters via the effective order parameter $\epsilon_n(g)$, implicitly determined by
\begin{equation}
\frac{\Gamma\!\left(\tfrac{1}{2}-\epsilon_n(g)\right)}{\Gamma(-\epsilon_n(g))} = -\tfrac{g}{2}.
\end{equation}
The normalization constant $A_n(g)$ is given by
\begin{equation}
A_{n}(g) = 2 \sqrt{\frac{\Gamma\!\left( \tfrac{1}{2} - \epsilon_n(g) \right)}{\Gamma(-\epsilon_n(g))}
     \frac{1}{\psi( \tfrac{1}{2} - \epsilon_n(g) ) - \psi(-\epsilon_n(g))}},
    \label{eqn:normalisation_constant}
\end{equation} 
with $\psi(z)=\Gamma'(z)/\Gamma(z)$ the digamma function.
The corresponding even-parity energies are $E_{2 n} = \hbar \omega \bigl(2 \epsilon_n(g) + 1/2\bigr)$.

A key analytically known property of these states is their two-body overlap \cite{Budewig2019},
\begin{equation}
        {C}_{2n;2m}(g_i, g_f)
        =\frac{A_{n}(g_i)A_{m}(g_f)}{g_ig_f}\,
        \frac{g_i-g_f}{E_{2n}(g_i)-E_{2m}(g_f)},
	\label{eqn:overlap_Laura_gi_gf}
\end{equation}
which can be used to compute overlap coefficients between the Busch basis and the noninteracting HO basis $U^{{\rm HO} \rightarrow {\rm B}}_{m, n}(g) = \braket{\psi_{m}^{\rm rel}|\psi_n^{\rm HO}}$ given by
\begin{equation}
U^{\mathrm{HO}\to\mathrm{B}}_{m,n}(g)=
\begin{cases}
\frac{1}{\sqrt[4]{\pi}}\,
\frac{A_{m/2}(g)}{E_m(g)-\tfrac12},
& n=0,\; m\ \text{even},\\[0.6em]
\frac{C_n}{\sqrt[4]{\pi}}\,
\frac{(-1)^{m/2}A_{m/2}(g)}{E_m(g)-\tfrac{2n+1}{2}},
& n\neq0,\; n,m\ \text{even},\\[0.6em]
\delta_{m,n}, & \text{otherwise}
\end{cases}
\label{eqn:Overlap_for_g0}
\end{equation}
with $C_n=\sqrt{\frac{(n-1)!}{2^{n-1}\left(\frac{n}{2}\right)!\left(\frac{n}{2}-1\right)!}}$.
These coefficients define the unitary maps
$U^{{\rm HO} \rightarrow {\rm B}}(g)$ and
$U^{{\rm B} \rightarrow {\rm HO}}(g) = [U^{{\rm HO} \rightarrow {\rm B}}(g)]^{\dagger}$,
which we use to switch between the two bases. % interacting Busch basis and the HO basis.
This transformation will later be essential for expressing the wavefunction in laboratory coordinates and in the HO momentum basis, which is required to analyze the Kapitza--Dirac diffraction pattern.

\subsection{Matrix Elements of the Hamilton operator}
\label{sec:MatrixElementsHamilton}
In the CM variables of Eq.~\eqref{eqn:coordinate_transformation}, the two-body Hilbert space factorizes as
$\mathcal H \cong \mathcal H_{\mathrm{CM}} \otimes \mathcal H_{\mathrm{rel}}$.
In the absence of the lattice term ($U_0=0$), the Hamiltonian separates as
$H_0=H_{\mathrm{CM}}+H_{\mathrm{rel}}$,
with eigenpairs
\begin{equation}
H_{\mathrm{CM}}\ket{\psi^{\mathrm{CM}}_n}=E^{\mathrm{CM}}_n\ket{\psi^{\mathrm{CM}}_n},
  \quad H_{\mathrm{rel}}\ket{\psi^{\mathrm{rel}}_k}=E^{\mathrm{rel}}_k\ket{\psi^{\mathrm{rel}}_k}.
\end{equation}
Hence, one has $E_n^{\rm CM}= \hbar \omega (n+\tfrac{1}{2})$.
Whereas in the relative sector, odd-parity states ($k=2n+1$) are unaffected by the contact interaction and we have $E_{2n+1}^{\rm rel}= \hbar \omega (2n+\tfrac{3}{2})$, while even-parity states ($k=2n$) are shifted to $E_{2n}^{\rm rel}= \hbar \omega \bigl(2\epsilon_n(g)+\tfrac{1}{2}\bigr)$.

We work in the orthonormal product basis
$\ket{n,k}\equiv\ket{\Psi^{\mathrm{HO}}_n}\otimes\ket{\Psi^{\mathrm{B}}_k}$,
for which
\begin{equation}
  \langle n,k|H_0|n',k'\rangle=(E^{\mathrm{HO}}_n+E^{\mathrm{B}}_k)\,\delta_{nn'}\delta_{kk'}.
\end{equation}
The lattice operator $ H_{\rm lat}$ depends on both $R$ and $r$ and therefore couples the center-of-mass and relative degrees of freedom. Therefore, the matrix elements of the total Hamilton operator are
\begin{align}
  \langle n,k|H|n',k'\rangle
  &=(E^{\mathrm{CM}}_n+E^{\mathrm{rel}}_k+U_0)\,\delta_{nn'}\delta_{kk'}\nonumber\\
  &\quad +\,U_0\,S^{\rm CM}_{n,n'}\,S^{\rm rel}_{k,k'}(g),
  \label{HO_matrix_elements_general}
\end{align}
where the couplings are generated by 
$\cos(\sqrt{2}\,k_{\rm lat}\, R)$ and
$\cos(\sqrt{2}\,k_{\rm lat}\, r)$, leading to 
\begin{align}
S^{\rm rel}_{n,m}(g)&=\big\langle \psi^{\rm rel}_{n}(g)\big|\cos(\sqrt{2}\,k_{\rm lat}\,r)\big|\psi^{\rm rel}_{m}(g)\big\rangle, \\
S_{n,m}^{\rm CM}&=S_{n,m}^{\rm HO}=\big\langle \psi^{\rm HO}_{n}\big|\cos(\sqrt{2}\,k_{\rm lat}\,R)\big|\psi^{\rm HO}_{m}\big\rangle.
\end{align}
By parity, $S^{\rm rel}_{n,m}(g)=0$, $S^{\rm CM}_{n,m}=0$ whenever $|n-m|$ is odd.
For odd relative states $2n+1$, $n = 0, 1, 2, \dots$, we obtain
\begin{align}
S_{2m+1,2n+1}^{\rm rel}
&= \big\langle \psi^{\rm HO}_{2m+1}\big|\cos(\sqrt{2}\,k_{\rm lat}\,r)\big|\psi^{\rm HO}_{2n+1}\big\rangle\nonumber\\
&=S_{2m+1,2n+1}^{\rm HO}.
\end{align}
For even relative states, we expand the Busch eigenfunctions in the HO basis using Eq.~\eqref{eqn:Overlap_for_g0} and employ the displacement-operator representation derived in Appendix~\ref{sec_app:Mat_Elemenets_and_Displacement_Operator}. This yields the unitary similarity transform
\begin{align}
S^{\rm rel}_{2m,2n}(g) &= \sum_{j,j'} U^{{\rm HO}\to{\rm B}}_{2m,j}(g)\,S^{\rm HO}_{j,j'}\,\left[U^{{\rm HO}\to{\rm B}}_{2n,j'}(g)\right]^*,
\end{align}
where $\left[U^{{\rm HO}\to{\rm B}}(g)\right]^{\dagger}$ maps even Busch states to HO states as derived in \eqref{eqn:Overlap_for_g0}.
The HO matrix elements admit the closed form
\begin{align}
S_{n,m}^{\mathrm{HO}} &=
e^{-(k_{\mathrm{lat}}\ell)^{2}/2} 
\sqrt{\frac{\min(n,m)!}{\max(n,m)!}} \,
(k_{\mathrm{lat}}\ell)^{\,|n-m|} \,
 \nonumber\\&\quad\times\,
L_{\min(n,m)}^{(|n-m|)}\!\left((k_{\mathrm{lat}}\ell)^{2}\right)\,\cos\!\Big(\tfrac{\pi}{2}\,|n-m|\Big),
\label{HO_lattice_matrix_elements_main}
\end{align}
where $L^{(a)}_{n}(x)$ are associated Laguerre polynomials \cite{gradshteyn_ryzhik_2014}.
The matrix $S^{\mathrm{HO}}$ is real, symmetric, and vanishes for $|n-m|$ odd.
In practice, we truncate to the finite subspace
$\mathrm{span}\{\ket{n,k}:\,0\le n\le m_{\rm max},\;0\le k\le m_{\rm max}\}$ and diagonalize the resulting Hermitian matrix.
The choice of $m_{\rm max}$ required for numerical convergence is analyzed in detail in Appendix~\ref{sec:conv_analysis}.
\subsection{Time evolution and observables}
\label{sec:time_evolution_observables}

To study Kapitza--Dirac scattering, we start from the ground state of the lattice-free Hamiltonian,
\begin{align}
  H_0 \ket{0,0}
  &=
  \bigl(E^{\rm HO}_0+E^{\rm B}_0\bigr)\ket{0,0},\nonumber\\
  \qquad
  \ket{0,0}&=\ket{\psi^{\rm CM}_0}\otimes\ket{\psi^{\rm rel}_0(g)},
  \label{eq:initial_state}
\end{align}
with $U_0=0$.
At $t=0$, we suddenly switch on the lattice to a constant depth $U_0$ (sudden quench) \cite{Idziaszek2006} and keep it on for a pulse duration $\tau$. For $0< t\leq \tau$ the Hamiltonian is
\begin{equation}
H=H_{\rm CM}+H_{\rm rel}+H_{\rm lat},
\end{equation}
and the time evolution obeys
\begin{equation}
  i\hbar\,\partial_t\ket{\Psi(t)}=H\ket{\Psi(t)}.
\end{equation}
Since $H$ is time-independent during the pulse, the formal solution is
\begin{equation}
  \ket{\Psi(t)}=\exp(-iH t/\hbar)\,\ket{0,0}.
  \label{eq:unitary_evolution}
\end{equation}
In practice, the time-dependent state $\ket{\Psi(t)}$ is obtained by propagating the initial state with Eq.~\eqref{HO_matrix_elements_general} in the truncated basis, using an adaptive fourth–fifth–order Runge–Kutta (RK45) integrator.

A short lattice pulse acts as a phase grating: it imprints a spatially periodic phase on the atoms without substantial motional evolution during the pulse. After the pulse, the wavepacket contains discrete momentum components separated by $\pm 2\hbar k_{\rm lat}$, characteristic of Kapitza--Dirac diffraction.
In the following, we construct the one- and two-body densities and the momentum distribution, which constitute the central observables in our analysis.

\subsection{Density and correlation functions}
\label{sec:density_correlation_function}
A crucial step in evaluating observables is to transform from the CM frame back to the laboratory coordinates $(x_\uparrow,x_\downarrow)$. This is implemented by an effective $\pi/2$ rotation, as detailed in Appendix~\ref{sec_app:Rotation_to_Lab_frame}, which yields the expansion
\begin{equation}
    	\Psi(x_{\uparrow}, x_{\downarrow}; t)
    	=\sum_{m,n}\kappa_{m,n}(t)\,\Psi_m^{\rm HO}(x_{\uparrow})\,\Psi_n^{\rm HO}(x_{\downarrow}),
        \label{eqn:one_body_HO}
\end{equation}
with time-dependent coefficients $\kappa_{m,n}(t)$ determined by the CM coefficients, see Appendix~\ref{sec_app:Rotation_to_Lab_frame}.

Hence, the one-body reduced density matrices are given by
\begin{subequations}
\begin{align}
  \rho^{(1)}_{\uparrow}(x,x';t)
  &= \int_{\mathbb R} dx_{\downarrow}\,\Psi(x,x_{\downarrow};t)\,\Psi^*(x',x_{\downarrow};t),
    \label{eqn:reduced_one_body_matrixa}
\\
  \rho^{(1)}_{\downarrow}(x,x';t) 
  &= \int_{\mathbb R} dx_{\uparrow}\,\Psi(x_{\uparrow},x;t)\,\Psi^*(x_{\uparrow},x';t),
    \label{eqn:reduced_one_body_matrixb}
\end{align}
\end{subequations}
and the corresponding one-body densities are the diagonal elements
\begin{subequations}
\begin{align}
  n_{\uparrow}(x,t)&=\rho^{(1)}_{\uparrow}(x,x;t),
\\
  n_{\downarrow}(x,t)&=\rho^{(1)}_{\downarrow}(x,x;t),
    \label{eqn:one_body_density}
\end{align}
\end{subequations}
each normalized as $\int dx\,n_\sigma(x,t)=1$ for $\sigma\in\{\uparrow,\downarrow\}$.

To go beyond the single-particle picture, we introduce the two-body density
\begin{align}
    \rho^{(2)}(x_{\uparrow},x_{\downarrow};t)
    =\big|\Psi(x_{\uparrow},x_{\downarrow};t)\big|^2,
    \label{eqn:two_body_density}
\end{align}
normalized such that $\int dx_\uparrow dx_\downarrow\,\rho^{(2)}(x_\uparrow, x_\downarrow; t)= 2$.
Its structure along and away from the diagonal $x_\uparrow=x_\downarrow$ directly reflects interaction-induced correlations: suppression along the diagonal for repulsive interactions and enhancement for attractive ones.
To focus on spatial correlations in the relative degree of freedom, we consider the projected relative density
\begin{equation}
\rho_{\rm rel}(s,t) = \int \mathrm{d}x\,\rho^{(2)}(x, x+s;t),
\label{eq:rel_density_continuous}
\end{equation}
which is large near $s=0$ for strongly attractive, tightly bound pairs and shifts to finite $|s|$ as repulsive interactions drive spatial separation.

On a numerical grid with $N$ points, the two-body density is represented as
$\rho^{(2)}_{i,j}(t) = \rho^{(2)}(x_i, x_j; t)$. The relative density can then be constructed by summing along diagonals of constant separation,
\begin{equation}
\rho_{\rm rel}(s, t)
= \sum_{j-i=s} \rho^{(2)}_{i,j}(t),
\qquad
s \in \{-N{+}1,\dots,N{-}1\},
\label{eq:diag_sum_two_body_density}
\end{equation}
which provides a compact measure of pair correlations: large weight near $s=0$ signals strong binding, whereas enhanced probability at finite $s$ reflects repulsion-induced spatial separation.

The experimentally accessible momentum distribution is obtained by Fourier transforming the one-body density matrices,
\begin{equation}
  n_\sigma(k,t)=\frac{1}{2\pi}\int_{\mathbb R}dx\,dx'\;e^{-ik(x-x')}\,\rho^{(1)}_\sigma(x,x';t),
\end{equation}
for $\sigma\in\{\uparrow,\downarrow\}$.
These distributions exhibit Kapitza--Dirac diffraction peaks at momenta separated by $\pm 2\hbar k_{\rm lat}$ and constitute the primary observable in a possible comparison with experiments.

Using the expansion in HO basis functions from Eq.~\eqref{eqn:one_body_HO}, the momentum-space HO eigenfunctions read
\begin{align}
\psi_n^{\rm HO}(k_{\sigma}) &= \frac{i^{3n}}{\sqrt{2^n n!}} \left( \frac{\ell^2}{\pi \hbar^2} \right)^{1/4}\nonumber\\&\quad \times H_n\left({ k_{\sigma}}{\ell}\right) \exp{\left[- \frac{(k_{\sigma}\ell)^2}{2}\right]},
\end{align}
where the factor $i^{3n}$ ensures phase consistency with the spatial representation.
In terms of the expansion coefficients $\kappa_{m,n}(t)$, the momentum densities can be expressed directly in the HO momentum basis as
\begin{subequations}
\begin{align}
  n^{(1)}_{\uparrow}(k,t)
  & = \sum_{m, m', n} \kappa_{m',n}^*(t)\,\kappa_{m,n}(t)\,
  \big[\Psi_{m'}^{\rm HO}(k)\big]^*\,\Psi_m^{\rm HO}(k),\\
  n^{(1)}_{\downarrow}(k,t) 
  & = \sum_{n, n', m} \kappa_{m,n'}^*(t)\,\kappa_{m,n}(t)\,
  \big[\Psi_{n'}^{\rm HO}(k)\big]^*\,\Psi_n^{\rm HO}(k).
    \label{eqn:reduced_one_body_matrix_momentum}
\end{align}
\end{subequations}

Having defined the relevant observables, we now contrast the full dynamics with the sudden approximation, which provides a simplified description of short lattice pulses.
\subsection{Sudden Approximation}
\label{sec:sudden_approximation_theory}
Kapitza--Dirac scattering with short lattice pulses can be described within the sudden (or delta-kick) approximation, widely used in atomic, molecular, and optical physics \cite{Moskalenko2017}. If the pulse duration $\tau$ is much shorter than the characteristic timescales of the system’s free evolution under $H_0$, the internal dynamics during the pulse can be neglected and the evolution is dominated by the lattice term $H_{\rm lat}$.

The exact time-evolution operator from $t_0$ to $t_f$ is
\begin{equation}
U(t_f,t_0) = \mathcal{T}\exp\left[-\frac{i}{\hbar}\int_{t_0}^{t_f}dt\big(H_0+U_0{O}\big)\right],
\label{eqn:exact_time_evolution_operator}
\end{equation}
where $\mathcal{T}$ denotes time ordering and ${O}=S^{\rm rel}(g)S^{\rm CM}$ is the operator that encodes the lattice-induced coupling of center-of-mass and relative motion \cite{Shankar1994}. In the sudden limit, we approximate the lattice by a delta-like pulse of fixed area
\begin{equation}
    K=\frac{1}{\hbar}\int_0^\tau U_0\,dt.
\end{equation}
In the interaction picture, the associated Magnus expansion \cite{Magnus1954} truncates after the first term, because the delta pulse collapses all commutators $[O_I(t_1),O_I(t_2)]$ to zero. The result is the factorization
\begin{equation}
U_{\rm sudden}(\tau) = e^{-i H_0(\tau-t_0)/\hbar}\,e^{-iK{O}}\,e^{-i H_0 t_0/\hbar}.
\end{equation}
In the basis of the $H_0$ eigenstates, see Sect.~\ref{sec:MatrixElementsHamilton}, the outer free-evolution operators acting before and after the delta kick, contribute global phase factors, $e^{-iH_0 t/\hbar}|\Psi_0\rangle = e^{-iE_0 t/\hbar}|\Psi_0\rangle$. Thus the physical action of the pulse within the sudden approximation is entirely contained in the kick operator
\begin{equation}
U_K(K) = e^{-iK O}.
\label{eqn:kick_operator}
\end{equation}
This operator imprints a spatially periodic phase on the wavefunction, i.e. it acts as a phase grating \cite{Clauser1994}. In our two-body setting, the diffraction acts on the interaction-dressed relative wavefunction \cite{Busch1998}, providing a sensitive probe of correlations.
\begin{figure*}
    \centering
    \includegraphics[width=1.\linewidth]{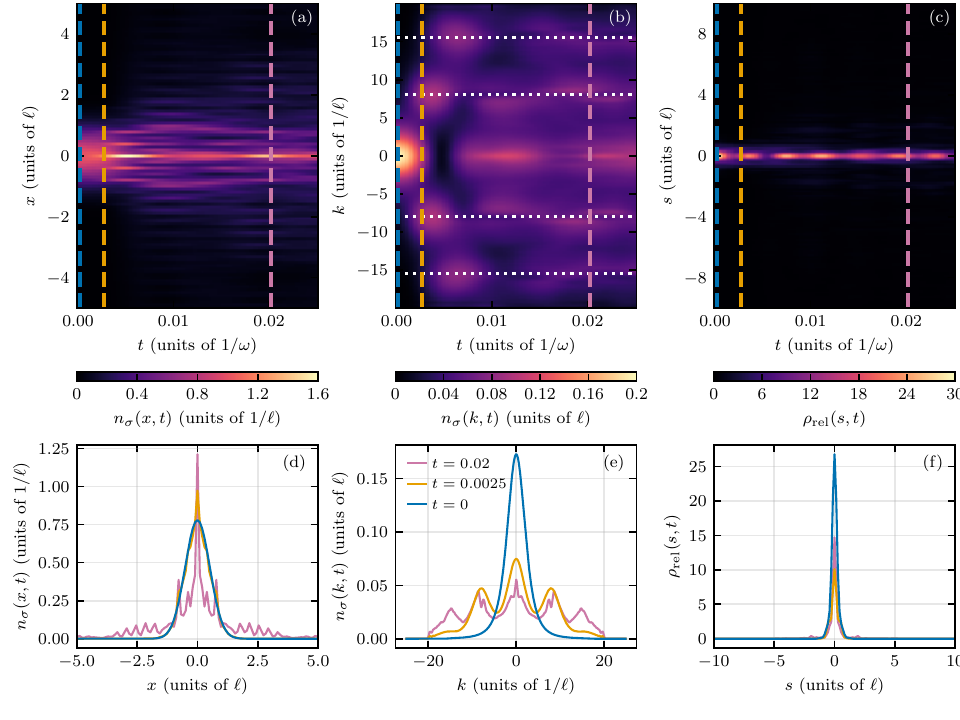}
    \caption{
(a) Time evolution of the one-body density of two interacting particles with interaction strength 
$ g = -5\hbar\omega\ell $ in a lattice with $ k_{\mathrm{lat}} = 4/\ell $ 
and lattice depth $ U_{0} = -1000\hbar\omega $.
(b) Corresponding momentum distribution; dotted lines indicate 
$ k = \pm 2k_{\mathrm{lat}} $ and $ k = \pm 4k_{\rm{lat}}$.
(c) Two-body correlation function.
In all panels, the dashed lines indicate the time slices at 
$ t = 0 $, 
$ t = 2.5\times10^{-3}/\omega $ and 
$ t = 2.0\times10^{-2}/\omega $, for which the corresponding line profiles are displayed in panels (d)--(f).
}
    \label{fig:g-5_k4_U1000}
\end{figure*}

To quantify the accuracy of the sudden approximation, we compare the corresponding state
\begin{equation}
    |\Psi_{\rm sudden}(\tau)\rangle=e^{-i{H}_0\tau/\hbar}\,U_K(K)\,|\Psi_0\rangle 
\end{equation}
with the exact time-evolved state
\begin{equation}
    |\Psi_{\rm full}(\tau)\rangle=U_{\rm full}(\tau)\,|\Psi_0\rangle 
\end{equation}
using the fidelity
\begin{equation}
F(\tau) = \big|\langle \psi_{\rm full}(\tau)|\psi_{\rm sudden}(\tau)\rangle\big|^2.
\label{eqn:Fidelity}
\end{equation}
For the short pulses typical of Kapitza--Dirac experiments, $F(\tau)$ is expected to remain close to unity, confirming the validity of the sudden approximation \cite{Gorin2006}. At longer times, deviations increase and demonstrate the breakdown of the impulsive regime, when the commutator $[{H}_0, {O}]$ can no longer be neglected. This crossover from sudden to non-impulsive dynamics has been investigated in both molecular physics and ultracold-atom contexts \cite{Ovchinnikov1999, Mueller2008, Reeves2015}.

With the theoretical framework established, we now turn to the dynamical results and the manifestation of Kapitza--Dirac scattering in the one- and two-body observables.
\section{Results}
\label{sec:Results}
\begin{figure*}
    \centering
    \includegraphics[width=1.\linewidth]{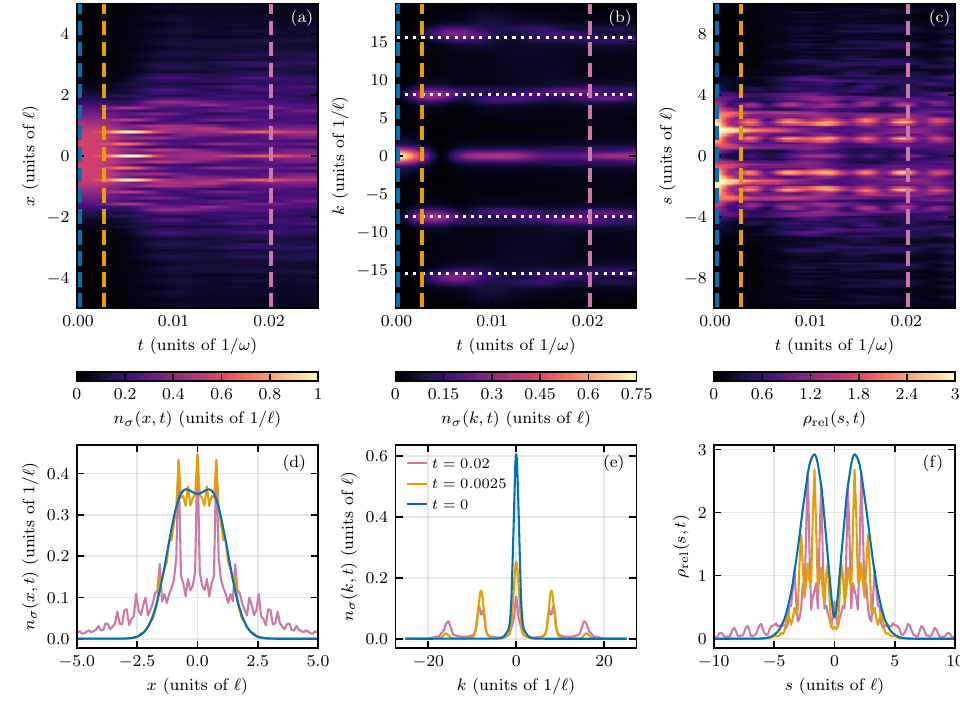}
        \caption{
(a) Time evolution of the one-body density of two interacting particles with interaction strength 
$ g = 5\hbar\omega\ell $ in a lattice with $ k_{\mathrm{lat}} = 4/\ell $ 
and lattice depth $ U_{0} = -1000\hbar\omega $.
(b) Corresponding momentum distribution; dotted lines indicate 
$ k = \pm 2k_{\mathrm{lat}} $ and $ k = \pm 4k_{\rm{lat}}$.
(c) Two-body correlation function.
In all panels, the dashed lines indicate the time slices at 
$ t = 0 $, 
$ t = 2.5\times10^{-3}/\omega $ and 
$ t = 2.0\times10^{-2}/\omega $, for which the corresponding line profiles are displayed in panels (d)--(f).
}
    \label{fig:g5_k4_U1000}
\end{figure*}
In this section we present the numerical results for different interaction strengths $g$, lattice depths $U_0$ and lattice wavevectors $k_{\rm lat}$. The choice of parameters is guided by the characteristic scales of state-of-the-art experiments \cite{Kiefer2023}. We analyze the one- and two-body densities, their corresponding momentum distributions, and the manifestation of Kapitza--Dirac scattering. 

\begin{figure*}
    \centering
    \includegraphics[width=1.\linewidth]{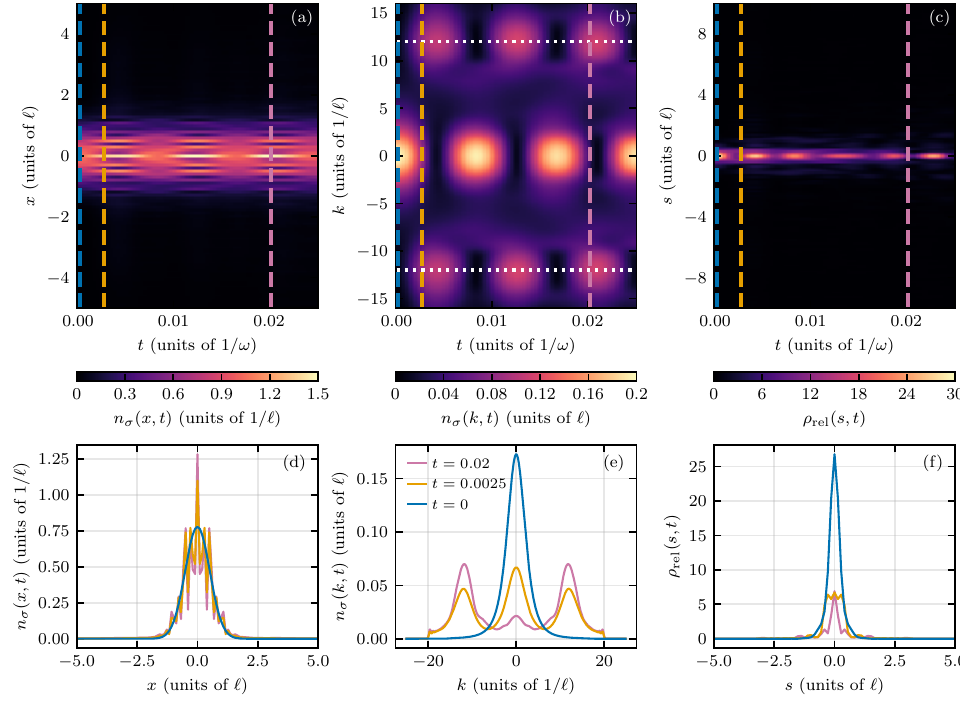}
 \caption{
(a) Time evolution of the one-body density of two interacting particles with interaction strength 
$ g =-5\hbar\omega\ell $ in a lattice with $ k_{\mathrm{lat}} = 6/\ell $ 
and lattice depth $ U_{0} = -1000\hbar\omega $.
(b) Corresponding momentum distribution; dotted lines indicate 
$ k = \pm 2k_{\mathrm{lat}} $.
(c) Two-body correlation function.
In all panels, the dashed lines indicate the time slices at 
$ t = 0 $, 
$ t = 2.5\times10^{-3}/\omega $ and 
$ t = 2.0\times10^{-2}/\omega $, for which the corresponding line profiles are displayed in panels (d)--(f).
}
    \label{fig:g-5_k6_U1000}
\end{figure*}

\begin{figure*}
    \centering
    \includegraphics[width=1.\linewidth]{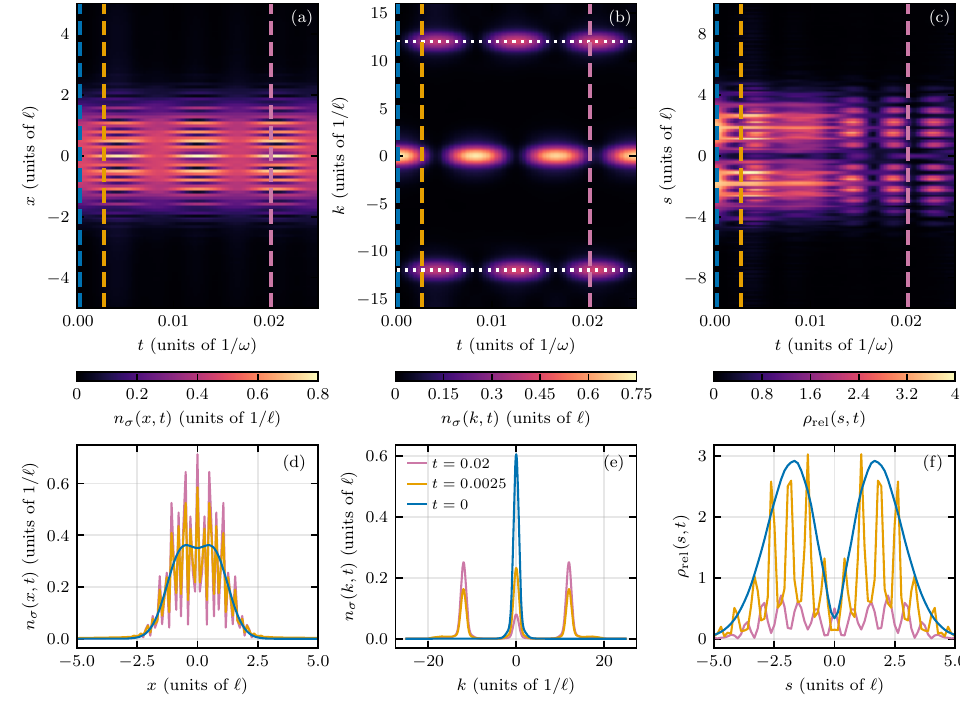}
 \caption{
(a) Time evolution of the one-body density of two interacting particles with interaction strength 
$ g =5\hbar\omega\ell $ in a lattice with $ k_{\mathrm{lat}} = 6/\ell $ 
and lattice depth $ U_{0} = -1000\hbar\omega $.
(b) Corresponding momentum distribution; dotted lines indicate 
$ k = \pm 2k_{\mathrm{lat}} $.
(c) Two-body correlation function.
In all panels, the dashed lines indicate the time slices at 
$ t = 0 $, 
$ t = 2.5\times10^{-3}/\omega $ and 
$ t = 2.0\times10^{-2}/\omega $, for which the corresponding line profiles are displayed in panels (d)--(f).
}
    \label{fig:g5_k6_U1000}
\end{figure*}

Before analyzing the dynamics induced by the lattice quench, we first discuss the structure of the interacting initial state in the harmonic trap. The one-body density, the momentum distribution, and the two-body correlation function at $t=0$ are shown in panels (a)–(c) of Figs.~\ref{fig:g-5_k4_U1000}–\ref{fig:g5_k6_U1000}, with the corresponding line profiles displayed in (d)–(f).

For strong attractive interactions ($g<0$), the one-body density develops a tightly localized peak around the trap center, reflecting the formation of an effectively bound pair, see Figs.~\ref{fig:g-5_k4_U1000} and \ref{fig:g-5_k6_U1000}(a,d). In contrast, strong repulsion ($g>0$) broadens the density significantly and produces two maxima displaced symmetrically from the origin, indicating strong spatial separation between the particles (see Figs.~\ref{fig:g5_k4_U1000}(a,d) and \ref{fig:g5_k6_U1000}(a,d)).

These spatial features directly imprint on the corresponding momentum distributions: strong attraction produces a broad momentum profile, while strong repulsion results in a narrower distribution concentrated near $k=0$. As defined in Eq.~\eqref{eq:diag_sum_two_body_density}, the projected relative density $\rho_{\rm rel}(s,t)$ provides a measure of spatial correlations between the two particles. For attractive interactions, $\rho_{\rm rel}(s,0)$ is sharply peaked around $s=0$, reflecting a high probability for the particles to occupy the same spatial region.
In contrast, repulsive interactions suppress weight near $s=0$ and shift probability toward finite relative separations, with symmetric maxima at $s\approx \pm 1.75\,\ell$, indicating interaction-induced spatial separation. These ground-state features provide the baseline against which the subsequent lattice-driven dynamics are interpreted.

Once the lattice is switched on at $t>0$, the one-body density develops oscillatory modulations dictated by the periodic lattice potential, while the momentum distribution shows emerging Kapitza--Dirac diffraction peaks at $k = 2 \nu k_{\rm lat}$, $\nu = \pm1, \pm 2, \dots$. We now analyze these dynamical features for the two different lattice wavevectors considered.

For $k_{\rm lat}=4/\ell$, the lattice spacing is relatively large and both attractive and repulsive interactions permit population of the first and second diffraction orders. In the attractive case (Fig.~\ref{fig:g-5_k4_U1000}), the one-body density remains largely localized in the $|x| < 2 \ell$ region and follows the lattice modulation, while the momentum distribution develops clear peaks at $\pm 2k_{\rm lat}$ and visible population at $\pm 4k_{\rm lat}$. For strong repulsion (Fig.~\ref{fig:g5_k4_U1000}), the spatial extent in the one-body density remains broad and the first-order peaks become particularly sharp; the second order also acquires finite population, demonstrating its energetic accessibility for this lattice momentum.

The two-body density retains the characteristic correlation structure set by the interaction: The projected relative density $\rho_{\rm rel}(s,t)$ retains the interaction-dominated correlation structure throughout the evolution. Attractive interactions keep the weight concentrated near small relative separations, while repulsive interactions maintain enhanced probability with symmetric maxima at $s\approx \pm 1.75\,\ell$, corresponding to spatial avoidance.
The lattice quench imprints a periodic temporal modulation on the relative-coordinate density over a broad interval of relative separations, with the dominant weight remaining concentrated in two symmetric peaks at $s \approx \pm 1.75\,\ell$. Therefore, the lattice does not qualitatively alter the interaction-induced correlation profile, but only modulates it in time. These modulations appear throughout the correlation plane, both near the trap center and in the outer regions, and follow the Kapitza--Dirac periodicity visible in the one-body observables.

For $k_{\rm lat}=6/\ell$, the recoil energy is significantly larger leading to a strong suppression of higher diffraction orders. As seen in Figs.~\ref{fig:g-5_k6_U1000} and \ref{fig:g5_k6_U1000}, only the first diffraction order at $\pm 2k_{\rm lat}$ becomes populated, while the second order remains completely absent due to the increased kinetic-energy cost,
\begin{equation}
\Delta E_n = \frac{(2 n \hbar k_{\rm lat})^2}{2m}.
\label{eqn:Energy_differences}
\end{equation}
However, the larger recoil energy enhances the coherence of the lattice-driven population transfer, such that the Kapitza--Dirac diffraction orders develop a distinctly periodic temporal pattern with sharply resolved maxima and minima in each order.
Interactions modulate the density in the same qualitative manner as in the $k_{\rm lat}=4/\ell$ case: attraction keeps the wave packet localized, whereas repulsion enhances the multi-peak structure across several lattice sites. Here, the interaction-driven correlation remains unchanged: The interaction-induced correlation pattern in $\rho_{\rm rel}(s,t)$ remains unchanged for the larger lattice momentum.
Attractive interactions preserve a dominant contribution at small relative separations, while repulsive interactions maintain weight at finite $|s|$. The increased recoil energy primarily suppresses higher-order lattice-induced features, without modifying the underlying two-body correlations or their temporal evolution.

\begin{figure*}
    \centering
    \includegraphics[width=1.\linewidth]{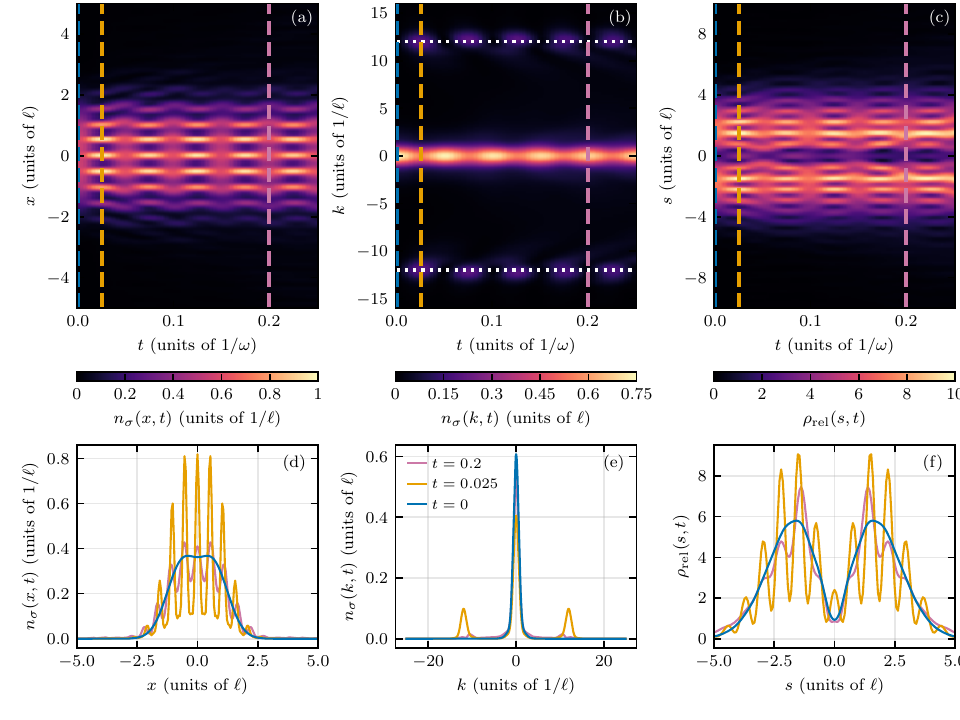}
 \caption{
(a) Time evolution of the one-body density of two interacting particles with interaction strength 
$ g =5\hbar\omega\ell $ in a lattice with $ k_{\mathrm{lat}} = 6/\ell $ 
and lattice depth $ U_{0} = -100\hbar\omega $.
(b) Corresponding momentum distribution; dotted lines indicate 
$ k = \pm 2k_{\mathrm{lat}} $.
(c) Two-body correlation function.
In all panels, the dashed lines indicate the time slices at 
$ t = 0 $, 
$ t = 2.5\times10^{-2}/\omega $ and 
$ t = 2.0\times10^{-1}/\omega $, for which the corresponding line profiles are displayed in panels (d)--(f).
}
    \label{fig:g5_k6_U100}
\end{figure*}
In a shallower lattice ($U_0 = -100\,\hbar\omega$), the features discussed above persist but are less pronounced. Here, the one-body density exhibits a periodic evolution between a Gaussian-like profile imposed by the harmonic confinement and a weakly lattice-modulated structure generated by the quenched lattice potential. The dynamics is therefore dominated by a smooth reshaping of the density rather than by pronounced localization within individual lattice wells. Owing to the reduced lattice depth, this evolution occurs on a slower timescale compared to the deeper-lattice case, as indicated by the corresponding time axes. Qualitatively, the behavior resembles that observed in Fig.~\ref{fig:g5_k6_U1000}(a), but the lattice-induced modulations are noticeably smoother, with fewer intermediate minima and maxima and reduced variations in peak height [see Figs.~\ref{fig:g5_k6_U1000}(d) and \ref{fig:g5_k6_U100}(d)].
The momentum distribution reflects this reduced modulation: as the Kapitza–Dirac peaks at $k = \pm 2 k_{\rm lat}$ possess much lower amplitude. This indicates that the shallow lattice does not impart sufficient momentum to populate higher scattering channels. Thus, the dynamics remains dominated by the central momentum component near $k=0$.
The corresponding two-body density exhibits only modest changes. The observed oscillations follow the periodic modulation of the one-body density but do not generate new correlation structures. The projected relative density $\rho_{\rm rel}(s,t)$ exhibits only weak temporal modulations and no qualitative redistribution in the relative coordinate. The probability remains concentrated at finite $|s|$, consistent with interaction-induced spatial avoidance. The temporal modulation closely follows the evolution of the one-body density.
\subsection{Interaction strength dependence}
\label{sec:interaction_strength_time_instant}
\begin{figure*}
    \centering
    \includegraphics[width=1.\linewidth]{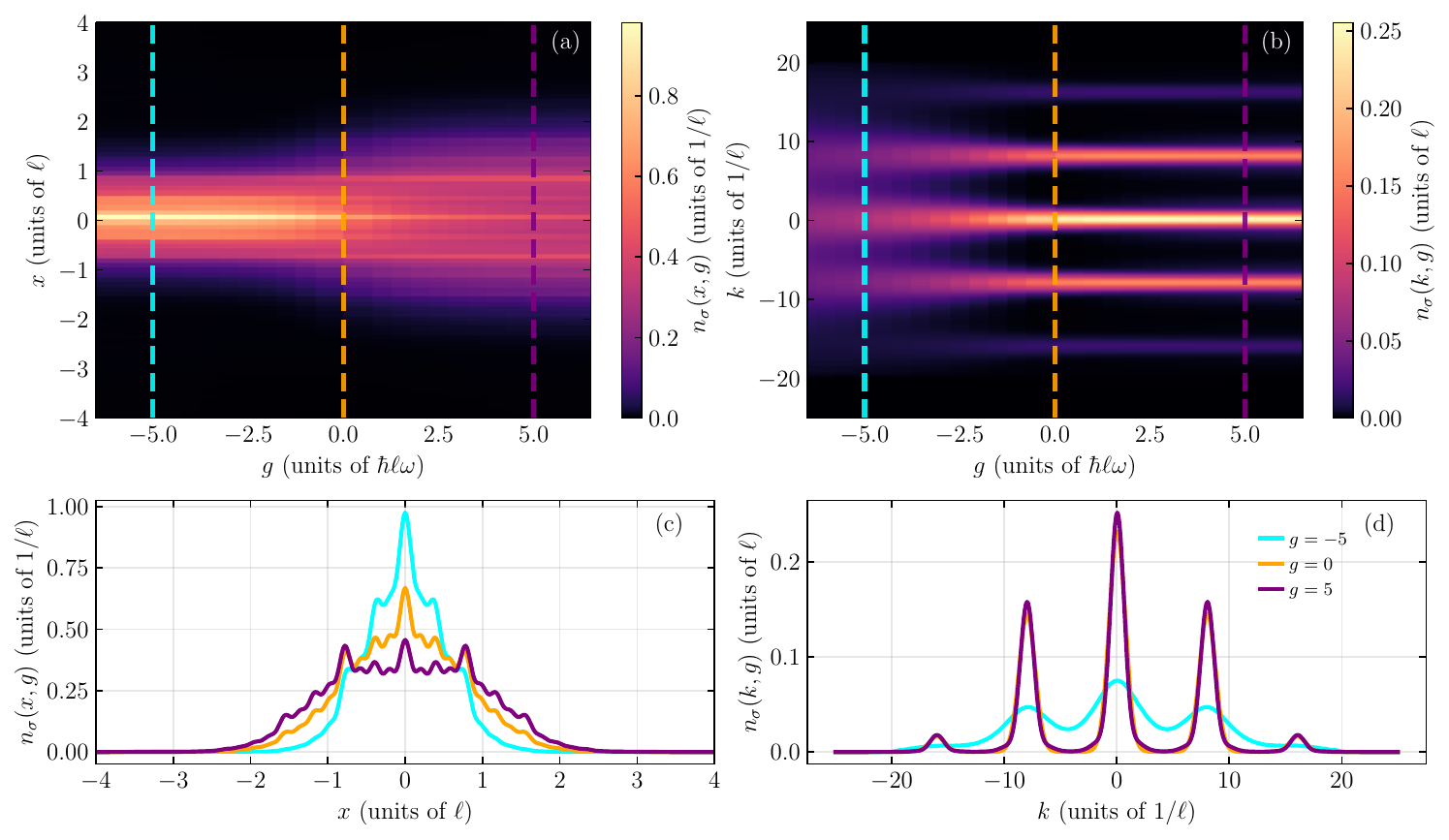}
    \caption{
(a) One-body density of two interacting particles at 
$ t_{0} = 2.5\times10^{-3}/\omega $ 
in a lattice with $ k_{\mathrm{lat}} = 4/\ell $ 
and $ U_{0} = -1000\hbar\omega $, 
shown as a function of the interaction strength $ g $.
(b) Corresponding momentum distribution. 
The colored dashed lines indicate the strongly attractive regime at 
$ g = -5\hbar\omega\ell $ 
and the noninteracting case $ g = 0 $ as well as the strongly repulsive interacting regime at $g=5\hbar\omega\ell$.
(c) and (d) show the corresponding line profiles of the 
one-body density and momentum distribution, respectively, 
for these interaction strengths.
}
    \label{fig:one_body_momentum_k4_U1000}
\end{figure*}
We now examine the time instant $ t_{0} = 2.5\times10^{-3}/\omega $ and compare the one-body density and momentum distribution as a function of the interaction strength $g$. Guided by the time evolution in Figs.~\ref{fig:g-5_k4_U1000} -- \ref{fig:g5_k6_U100}, we focus on this instant, lying close to where the first Kapitza--Dirac maximum is clearly developed in all considered cases. Throughout this section, we use the deep lattice $ U_{0} = -1000\,\hbar\omega $ to ensure a pronounced diffraction pattern, beginning with the case $ k_{\rm lat}=4/\ell $ as shown in Fig.~\ref{fig:one_body_momentum_k4_U1000}.

For strong repulsive interactions, the one-body density broadens as a result of interparticle repulsion and populates several lattice wells. The corresponding momentum distribution shows the opposite trend: the central peak and the first-order side peaks at $ k=\pm 2k_{\rm lat} $ become sharper and increase in contrast. For attractive interactions, the density narrows, while the momentum distribution broadens accordingly. As the interaction strength is varied, the density evolves from a sharply peaked profile for strong attraction to a broader distribution in the repulsive regime, approaching a lattice-modified Gaussian form. The diffraction response follows the same qualitative trends: attractive interactions smear the Kapitza--Dirac peaks, while the noninteracting and repulsive cases exhibit progressively clearer peak structure. Strong attractive interactions enhance real-space localization, which broadens the momentum distribution: the central and first-order diffraction peaks both broaden and flatten, while the second-order peaks at $k=\pm 4k_{\rm lat}$ become strongly suppressed in the strongly attractive regime [see Fig.~\ref{fig:one_body_momentum_k4_U1000} (d)]. However, the positions of the peaks remain fixed.

The physical mechanism underlying the sharpening of momentum peaks in the repulsive regime is twofold. First, the reduced spatial overlap of the particles suppresses the initial momentum spread, consistent with the distributions shown in Fig.~\ref{fig:g5_k4_U1000}. Second, the repulsive interaction suppresses density gradients, which reduces interaction-induced phase distortions from the quenched lattice potential, yielding a cleaner diffraction pattern. In the short-time regime considered here, the observed differences originate from the interaction-dependent structure of the initial state onto which the lattice phase is imprinted. Strong attractive interactions maximize spatial overlap and produce a tightly localized two-particle state. Acting on such a localized density, the lattice kick yields Kapitza–Dirac diffraction peaks that are broader and flatter. For weaker or repulsive interactions, the initially more extended density leads to narrower and more pronounced diffraction peaks.

\begin{figure*}
    \centering
    \includegraphics[width=1.\linewidth]{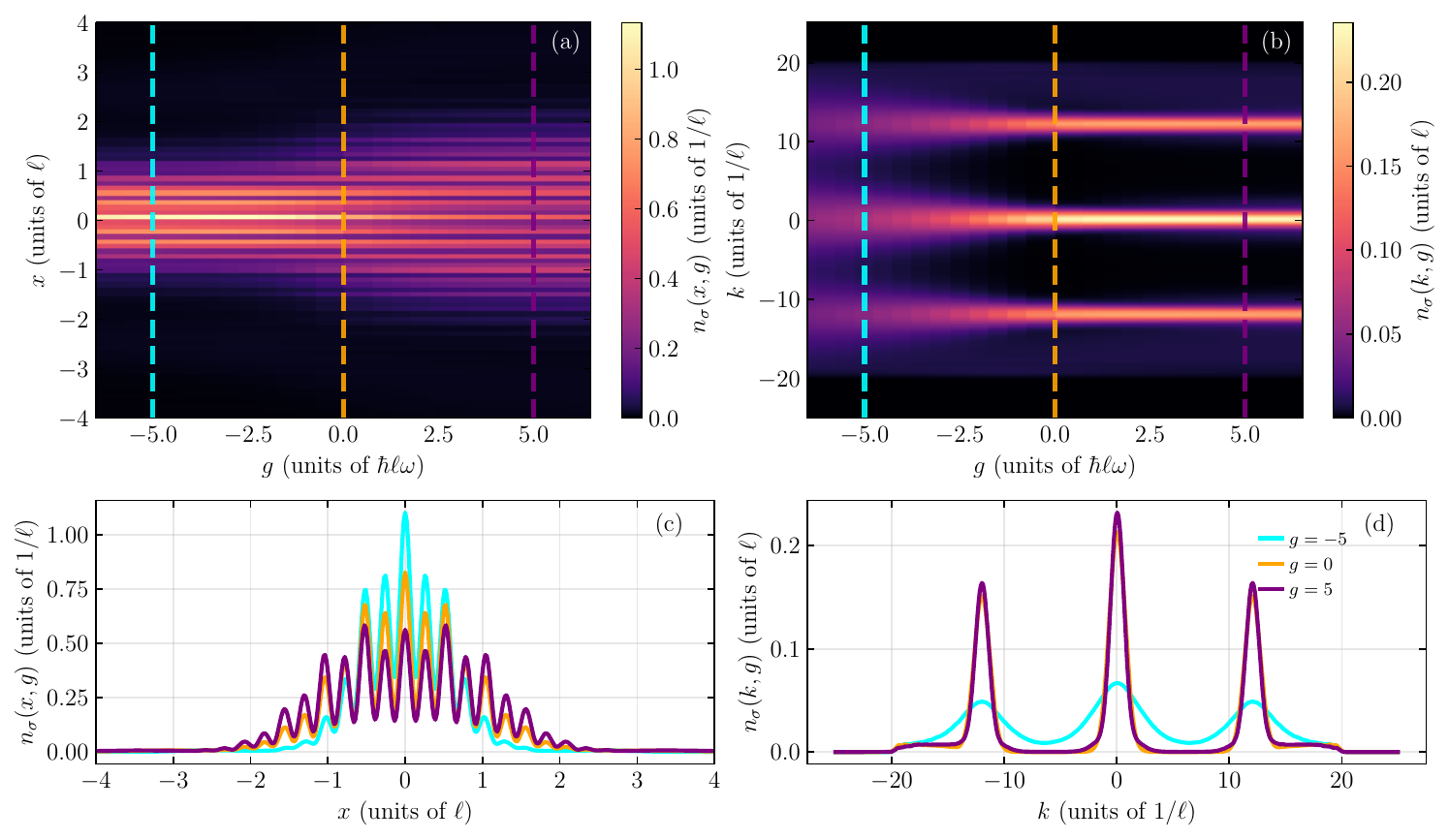}
\caption{
(a) One-body density of two interacting particles at 
$ t_{0} = 2.5\times10^{-3}/\omega $ 
in a lattice with $ k_{\mathrm{lat}} = 6/\ell $ 
and $ U_{0} = -1000\hbar\omega $, 
shown as a function of the interaction strength $ g $.
(b) Corresponding momentum distribution. 
Colored dashed lines indicate the strongly attractive regime at 
$ g = -5\hbar\omega\ell $ 
and the noninteracting case $ g = 0 $ as well as the strongly repulsive case $g=5\hbar\omega\ell$.
(c) and (d) show the corresponding line profiles of the 
one-body density and momentum distribution, respectively, 
for these interaction strengths.
}
    \label{fig:one_body_momentum_k6_U1000}
\end{figure*}
For $k_{\rm lat}=6/\ell$, the behavior changes characteristically compared with the case $k_{\rm lat}=4/\ell$. In the one-body density, the maxima move closer together due to the reduced lattice spacing, while the overall spatial extent of the density remains essentially unchanged. Unlike the three-peak structure observed for $k_{\rm lat}=4/\ell$, the larger lattice momentum produces a clear multi-peak pattern. Because the wells are more closely spaced, these smaller maxima appear relatively enhanced and form a more regular oscillatory structure across the central region.
The momentum distribution exhibits the complementary behavior. The central peak at k=0 remains comparable in height to the $k_{\rm lat}=4/\ell$ case, and the first-order diffraction peaks at $k=\pm 2k_{\rm lat}$ are still visible for all interaction strengths. However, the second-order components at $k=\pm 4k_{\rm lat}$ are entirely absent. This suppression follows directly from the recoil-energy requirement for Kapitza–Dirac scattering, as shown in Fig.~\ref{fig:g-5_k6_U1000} and Eq.\eqref{eqn:Energy_differences}: a larger lattice momentum increases the kinetic-energy spacing between diffraction orders, making population transfer into higher-order modes energetically inaccessible on the timescales considered. As a result, only the $n=1$ scattering order is populated, while the $n=2$ order remains unoccupied for all interaction strengths.
Despite these quantitative differences, the qualitative interaction dependence persists: attractive interactions narrow the one-body density and broaden the momentum distribution, whereas repulsive interactions broaden the density and sharpen the first-order diffraction peaks, consistent with the behavior observed for $k_{\rm lat}=4/\ell$ and summarized in Fig.~\ref{fig:one_body_momentum_k6_U1000}.

To disentangle the role of the lattice potential from the full dynamical evolution, we now examine the sudden approximation, where only the lattice term acts and the kinetic contribution is neglected. This allows us to examine how the initial state is redistributed purely by the lattice quench without considering any subsequent time evolution. 
\subsection{Sudden Approximation}
\label{sec:sudden}
\begin{figure*}
    \centering
    \includegraphics[width=1.\linewidth]{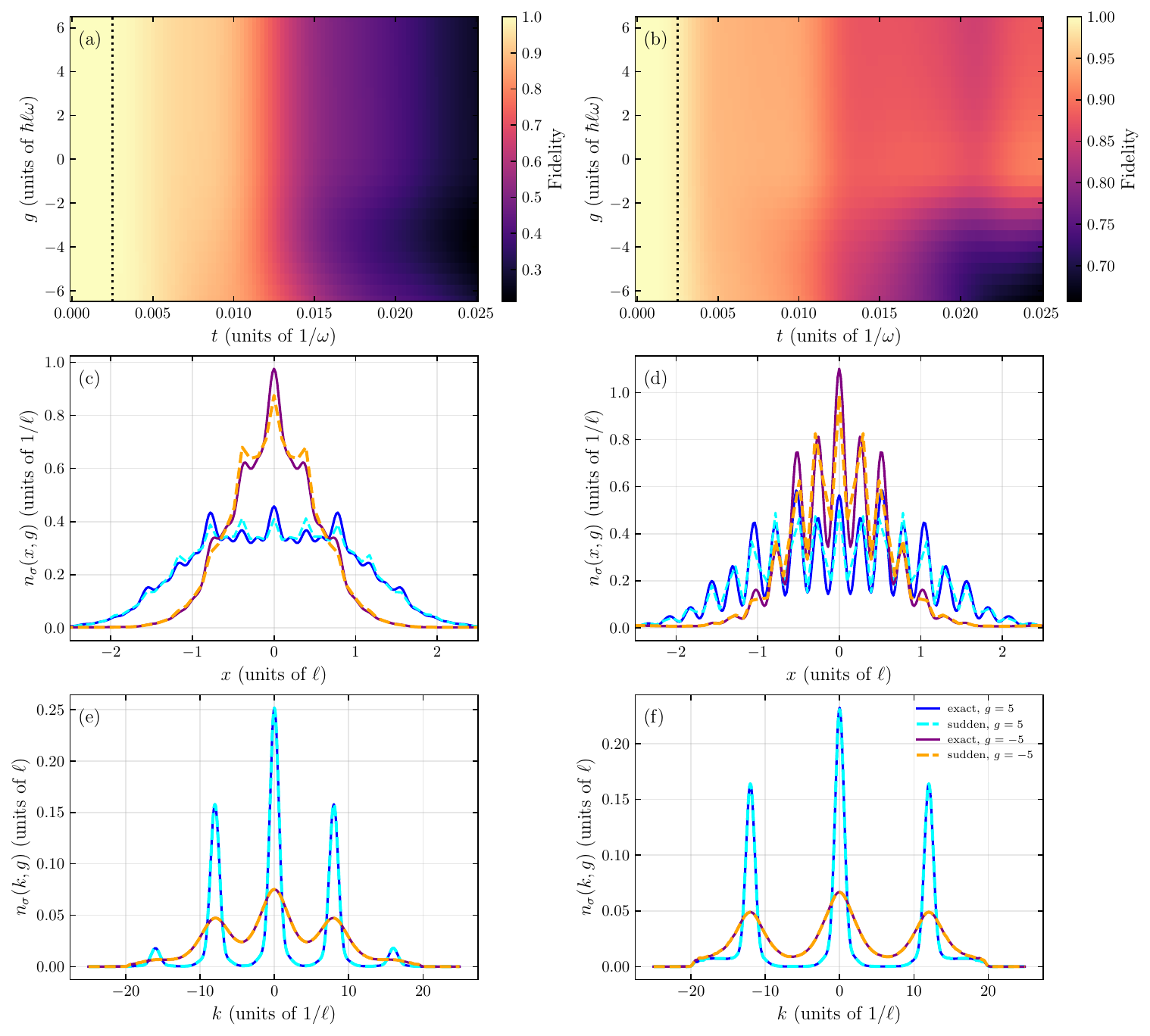}
    \caption{Fidelity of two interacting particles, comparing the overlap between the sudden approximation and the numerically exact results for $U_0 = -1000\hbar\omega$, shown for $k_{\rm lat} = 4/\ell$ in (a) and $k_{\rm lat} = 6/\ell$ in (b). The black dotted lines indicate the time snapshot at $t_0 = 2.5 \times 10^{-3}$. The corresponding one-body densities are displayed in (c) and (d), while the momentum distribution is shown in (e) and (f) at this time. These line plots allow for a direct comparison between the exact and sudden-approximation results for strongly attractive $(g = -5\,\hbar\omega\ell)$ and repulsive $(g = 5\,\hbar\omega\ell)$ interaction strengths.}
    \label{fig:fidelity_k4_k6}
\end{figure*}
In the following, we revisit the one-body density and momentum distribution for both $k_{\rm lat}=4/\ell$ and $k_{\rm lat}=6/\ell$ within the sudden-approximation framework and compare these predictions directly with the exact results for the one-body density and momentum distribution shown in Figs.~\ref{fig:one_body_momentum_k4_U1000} and \ref{fig:one_body_momentum_k6_U1000}. However, we first examine the fidelity, Eq.~\eqref{eqn:Fidelity}, in Figs.~\ref{fig:fidelity_k4_k6}(a) and (b), which provides a more sensitive measure of how accurately the sudden approximation reproduces the exact dynamics.

For both lattice momenta, the fidelity is initially close to unity for all interaction strengths, demonstrating that the sudden approximation captures the very early-time dynamics with high accuracy. At the time $t_{0}=2.5\times10^{-3}/\omega$ analyzed in Sect.~\ref{sec:interaction_strength_time_instant} (see dotted lines in Figs.~\ref{fig:fidelity_k4_k6}(a) and (b)), the overlap is essentially perfect for both $k_{\rm lat}=4/\ell$ and $k_{\rm lat}=6/\ell$. Thus, the qualitative features of the one-body density and momentum distribution remain nearly unchanged at short times, and we focus on direct line-plot comparisons for representative attractive ($g=-5\hbar\omega\ell$) and repulsive ($g=5\hbar\omega \ell$) interaction strengths in Figs.~\ref{fig:fidelity_k4_k6}(c)--(f). As time evolves, the dynamical effects generated by the kinetic energy, harmonic confinement, and interactions become increasingly important. In the sudden approximation, the dynamics generated by these contributions are assumed to be frozen during the instantaneous application of the lattice pulse, so that any nontrivial action of the initial Hamiltonian $H_0$ during the pulse is neglected and the full time evolution under $H_0$ resumes only after the delta-kick–modified state has been established. The lattice pulse imprints a spatially varying phase that is fully captured by the sudden approximation and directly shapes the momentum distribution, which therefore remains in very good agreement with the exact dynamics at short times [see Fig.~\ref{fig:fidelity_k4_k6} (e) and (f)].
First deviations appear in the real-space one-body density, where the imposed phase gradients begin to drive a weak spatial redistribution of the density during the pulse.
The fidelity, being sensitive to the full wavefunction including both density and phase information, amplifies these effects and thus provides the most sensitive indicator of the onset of deviations from the sudden approximation. Once these neglected contributions compete with the lattice-induced modulation, the two descriptions diverge and the fidelity decreases accordingly. For both lattice momenta, this becomes visible around $t \approx 1.0\times10^{-2}/\omega$, where the fidelity drops to values below $0.9$.

Beyond this point, the decay rate differs markedly between the two lattice momenta. For the smaller lattice momentum $k_{\rm lat}=4/\ell$, the fidelity then decreases rapidly and almost monotonically, reaching very low values for $t \gtrsim 2.0\times10^{-2}/\omega$. This signals a complete breakdown of the sudden approximation in this regime, particularly for both strongly attractive and strongly repulsive interactions, where the exact dynamics is strongly influenced by kinetic expansion and interaction-induced restructuring of the density. In contrast, for the larger lattice momentum $k_{\rm lat}=6/\ell$, the fidelity stabilizes around $\sim 0.8$ over the same time window for the non-interacting and repulsive cases, whereas it decreases to $\sim 0.7$ in the strongly attractive regime.
Here, the effectively stronger lattice action, arising from the larger recoil energy associated with the higher lattice momentum, keeps the dynamics more strongly dominated by the lattice modulation, so that neglecting kinetic, harmonic, and interaction terms is less decisive, and the sudden approximation remains a reasonably good description even at intermediate times.

The dependence on the interaction strength $g$ is comparatively weak but follows a clear trend. For $k_{\rm lat}=6/\ell$, the strongly attractive regime exhibits a markedly lower fidelity than the non-interacting limit, while the strongly repulsive case shows a more moderate reduction. This reflects the fact that attractive interactions enhance the local density in the center, as seen, e.g., in the direct comparison of the densities in Figs.~\ref{fig:fidelity_k4_k6}(c) and (d) and in the correlations in the exact evolution, thereby amplifying the role of the neglected interaction and kinetic contributions. Repulsive interactions also drive deviations, but to a lesser extent. 

This can already be seen—although only weakly—in the early-time snapshots of the one-body density and momentum distribution shown in Figs.~\ref{fig:fidelity_k4_k6}(c)–(f). At this short time, the deviations between the exact and sudden results are primarily visible in the one-body density, where kinetic- and interaction-induced dephasing begin to produce a slight reshaping of the spatial profile. By contrast, the momentum distribution still matches the sudden approximation almost perfectly, reflecting that diffraction features have not yet been significantly altered. These small but emerging real-space differences nevertheless foreshadow the rapid divergence observed at later times.
The strong fidelity decay for $k_{\rm lat}=4/\ell$ originates from dynamical effects that remain largely hidden in momentum space but are pronounced in real space. First indications of these deviations can already be seen in the early-time snapshots of the one-body density and momentum distribution shown in Figs.~\ref{fig:fidelity_k4_k6}(c)--(f). In the exact evolution, kinetic- and interaction-induced dephasing, together with a progressive accumulation of density inside the lattice wells, initiates a noticeable reshaping of the wavefunction. In contrast, the sudden approximation, acting purely as a phase grating, preserves a density profile set by the lattice modulation, causing the two evolutions to drift apart at longer times. Although the corresponding differences in the momentum distribution remain rather subtle, the real-space deviations dominate the overlap and are therefore responsible for the fidelity breakdown.

Overall, the sudden approximation provides an excellent description for short lattice pulses, where the dynamics is dominated by the instantaneous lattice impact. This matches the conditions of the Raman--Nath regime, in which a short lattice pulse acts primarily as a phase grating. However, at longer times the approximation breaks down. This breakdown is particularly pronounced for the smaller lattice momentum $k_{\rm lat}=4/\ell$, where kinetic spreading and interaction effects quickly generate deviations from the purely lattice-driven evolution. For larger $k_{\rm lat}$, the sudden approximation remains accurate for a longer interval, but ultimately the same limitations appear. Thus, while highly reliable for short-pulse Kapitza--Dirac scattering, the sudden approximation cannot capture the full dynamics at extended times, especially for small lattice momenta.

\section{Summary and Outlook}
\label{sec:Summary_Outlook}
We have theoretically investigated Kapitza--Dirac scattering in an interacting two-body system confined in a harmonic trap. Starting from the analytical solution of the two-body problem, we incorporated a lattice potential into the Hamiltonian and solved the time-dependent Schr\"odinger equation. From these solutions, we obtained the spatial one- and two-body densities as well as the corresponding momentum distributions, allowing for direct comparison with in situ and time-of-flight observables in current experiments.

Our results show that even a minimal two-body system captures the essential characteristics of Kapitza--Dirac scattering. After the lattice is turned on, discrete momentum components appear at $ k = \pm 2 k_{\rm lat} $, consistent with the expected diffraction orders. For smaller lattice momenta, higher-order diffraction peaks can also be populated. By varying the interaction strength, lattice depth, and lattice spacing, we systematically explored their influence on the scattering response. Attractive interactions lead to strong spatial localization and correspondingly broad momentum distributions, whereas repulsive interactions produce extended real-space densities and sharper momentum peaks. The two-body density reflects these trends through enhanced diagonal correlations for attractive interactions and increased anti-diagonal weight for repulsive interactions.

To further isolate the role of the lattice potential, we compared the exact dynamics with the sudden approximation, in which the lattice acts as a phase gradient that gets instantaneously imprinted in the two-body wavefunction while kinetic, interaction, and confinement contributions are neglected. This approximation captures the qualitative short-time diffraction dynamics but does not reproduce the full scattering behavior. The breakdown is most pronounced for small lattice momenta and at long times, particularly in the strongly attractive interaction regime. This comparison highlights that all terms of the Hamiltonian, including the lattice, kinetic, harmonic confinement, and interactions, contribute to the complete Kapitza--Dirac dynamics.

Overall, our study demonstrates that an exactly solvable two-body system already reproduces the key interaction-dependent features of Kapitza--Dirac scattering and provides a controlled benchmark for interpreting experiments in strongly correlated regimes. Although the present work focuses on one dimension, the underlying framework can be extended to higher dimensions. Using suitable unitary transformations, the two-body problem with a one-dimensional lattice can also be represented in two or three dimensions \cite{Budewig2019,Bougas2019}, enabling the exploration of e.g. fermionic pairing by examining its impact on Kapitza--Dirac scattering \cite{Cooper1956, Greiner2003, Zwierlein2003, Bayha2020}. We expect that the present results will serve as a benchmark for future many-body studies of interaction-driven diffraction dynamics.

\section*{Acknowledgements}
This work has been funded by the Cluster of Excellence “Advanced Imaging of Matter” of the Deutsche
Forschungsgemeinschaft (DFG) - EXC 2056 - project ID 390715994. G.K.M. has received funding by the Austrian Science Fund (FWF) [DOI: 10.55776/F1004].
\appendix 
\section{Convergence Analysis for the Busch Solutions}
\label{sec:conv_analysis}

\begin{figure}
    \centering
    \includegraphics[width=\linewidth]{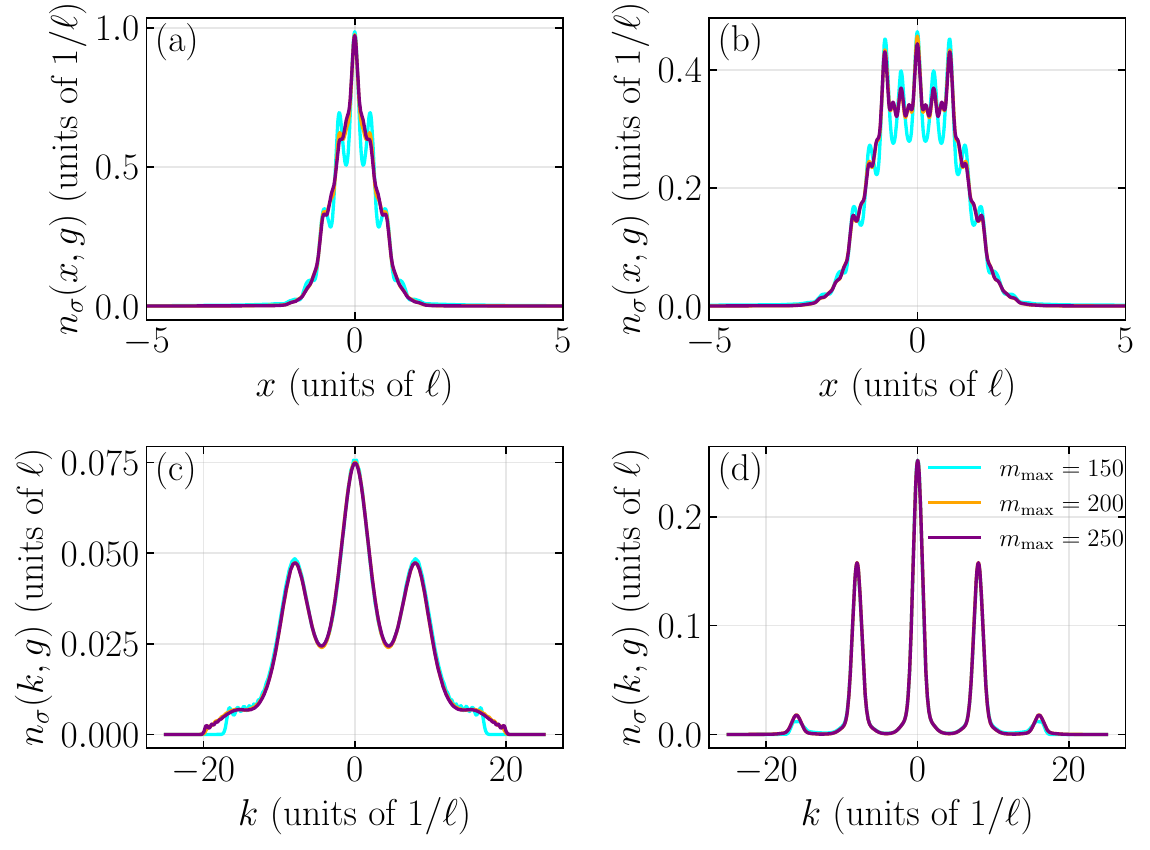}
    \caption{One-body density (a),(b) and momentum distribution (c),(d) for attractive 
    ($ g = -5\,\hbar\omega\ell $, left panels) and repulsive ($ g = 5\,\hbar\omega\ell $, right panels) 
    interaction strengths. In each panel, the different curves correspond to different values of 
    $ m_{\rm max} $ used in the initial truncation. All results are shown at the time snapshot 
    $ t_{0} = 2.5\times 10^{-3}/\omega $ for a lattice momentum $ k_{\rm lat} = 4/\ell $ and 
    lattice depth $ U_{0} = -1000\,\hbar\omega $.}
    \label{fig:one_body_momentum_k4_conv_analysis}
\end{figure}
In Sect.~\ref{sec:MatrixElementsHamilton} we introduced a truncation of the full two-body Hilbert space of the form $\mathrm{span}\{\ket{n,k}:\,0\le n\le m_{\rm max},\;0\le k\le m_{\rm max}\}$. Therefore, it is essential to demonstrate that the chosen truncation parameter $ m_{\rm max} $ is sufficient to capture all relevant physical features and that our results are not affected by cutoff artifacts.

\begin{figure}
    \centering
    \includegraphics[width=\linewidth]{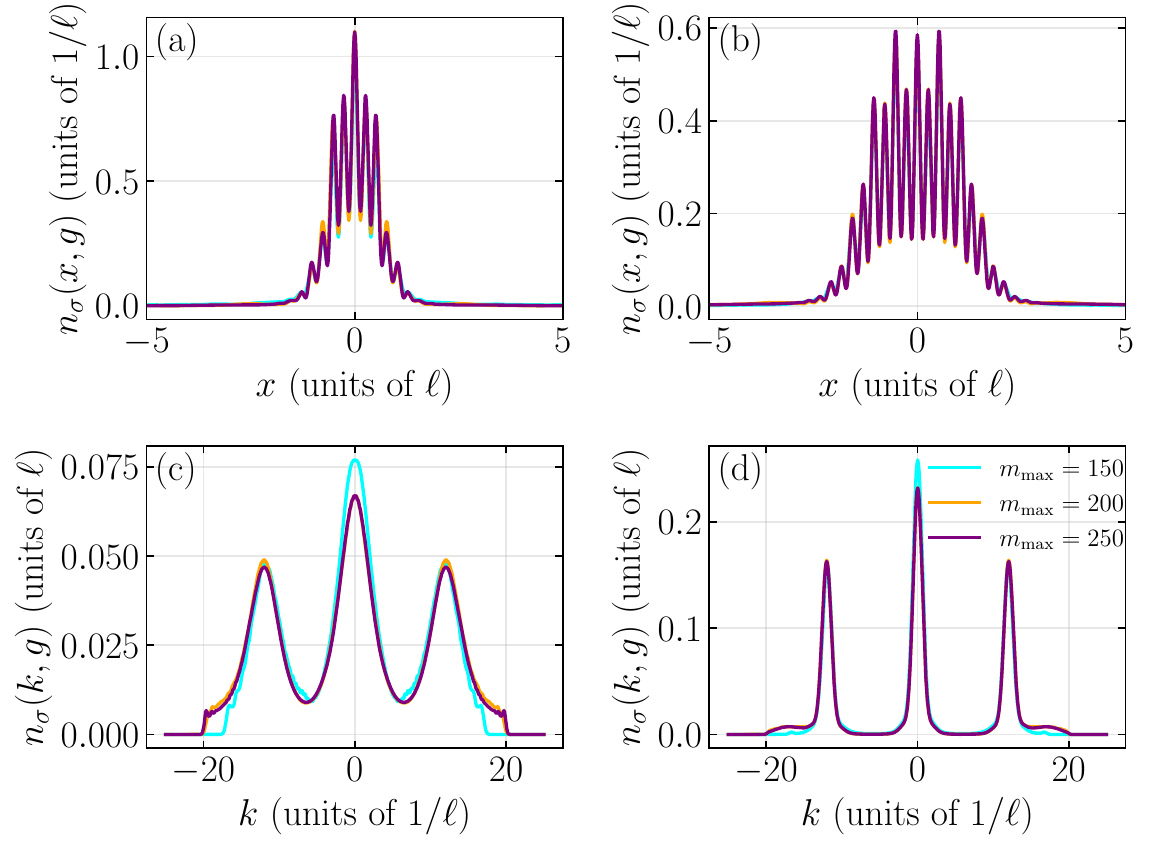}
    \caption{One-body density (a),(b) and momentum distribution (c),(d) for repulsive 
    ($ g = -5\,\hbar\omega\ell $, left panels) and attractive ($ g = 5\,\hbar\omega\ell $, right panels) 
    interaction strengths. In each panel, the different curves correspond to different values of 
    $ m_{\rm max} $ used in the initial truncation. All results are shown at the time snapshot 
    $ t_{0} = 2.5\times 10^{-3}/\omega $ for a lattice momentum $ k_{\rm lat} = 6/\ell $ and 
    lattice depth $ U_{0} = -1000\,\hbar\omega $.}
    \label{fig:one_body_momentum_k6_conv_analysis}
\end{figure}

Hence, in Figs.~\ref{fig:one_body_momentum_k4_conv_analysis} and 
\ref{fig:one_body_momentum_k6_conv_analysis} we focus on the convergence behavior of the most demanding parameter regime that occurs in our analysis. This regime corresponds to the deep lattice $ U_{0}=-1000\,\hbar\omega $, for which the higher-order Fourier components of the lattice potential contribute strongly, and to the strongest attractive and repulsive interactions $ g=\pm 5\,\hbar\omega\ell $. Furthermore, we focus on the characteristic time snapshot $t_0=2.5\times10^{-3}/\omega$, pointed out in Figs.~\ref{fig:one_body_momentum_k4_U1000} and \ref{fig:one_body_momentum_k6_U1000} to analyze the Kapitza--Dirac scattering. We investigate two lattice momenta, $ k_{\rm lat}=4/\ell $ and $ k_{\rm lat}=6/\ell $, to highlight the influence of the lattice spacing on convergence.

For $k_{\rm lat}=4/\ell$, the convergence behavior differs between the real and the momentum space. In the one-body density [see Fig.~\ref{fig:one_body_momentum_k4_conv_analysis}(a,b)], deviations in absolute values are visible for $m_{\rm max}=150$, while the positions of the minima and maxima are already well reproduced. Increasing the truncation to $m_{\rm max}=200$ yields quantitative agreement with the $m_{\rm max}=250$ results for both attractive and repulsive interactions. In momentum space, convergence is reached more rapidly: already for $m_{\rm max}=150$, the distributions closely match the converged result, with only minor deviations at large $|k|$, which vanish upon increasing $m_{\rm max}$.

For the larger lattice wavevector $k_{\rm lat}=6/\ell$, convergence becomes more demanding. In the one-body density [see Fig.~\ref{fig:one_body_momentum_k6_conv_analysis}(a,b)], only small differences remain between $m_{\rm max}=150$ and $m_{\rm max}=250$, reflecting the finer spatial oscillations. In contrast, the momentum distribution converges more slowly, particularly in the attractive regime, where peak shapes and steepness still depend on the truncation. For repulsive interactions, convergence is faster, with residual deviations mainly affecting the peak heights. Overall, increasing $k_{\rm lat}$ significantly enhances the numerical requirements to achieve convergence, especially in the momentum space.

In general, the convergence analysis confirms the validity and stability of the results presented in Sect.~\ref{sec:Results}. At the same time, it highlights the numerical limitations that arise for large lattice momenta and deep lattice depths, where the required basis size grows substantially due to the increasing contributions of higher-order lattice harmonics.

\section{Matrix Elements and Displacement Operator}
\label{sec_app:Mat_Elemenets_and_Displacement_Operator}
The analytic expression for the lattice term within the harmonic oscillator basis $S_{m,n}^{\rm HO}$ is evaluated with the help of the displacement operator. Below we summarize the derivation of this term.

The displacement operator $ \hat{D}(\alpha) $ generates coherent states by displacing the vacuum state in the phase space \cite{Loudon2000}. It is defined by
\begin{equation}
\hat{D}(\alpha) = e^{\alpha \hat{a}^\dagger - \alpha^* \hat{a}},
\end{equation}
where $ \alpha \in \mathbb{C} $ is a complex displacement and
$ \hat{a} $ and $ \hat{a}^\dagger $ are the  annihilation and creation operators.
Acting on the vacuum
\begin{equation}
|\alpha\rangle = \hat{D}(\alpha) |0\rangle
\end{equation}
which is a coherent state centered at $\alpha$ in phase space \cite{Glauber1963}. Using the Baker–Campbell–Hausdorff identity one also has the ordered form
\begin{equation}
\hat D(\alpha) \;=\; e^{-|\alpha|^2/2}\, e^{\alpha \hat a^\dagger}\, e^{-\alpha^* \hat a}.
\label{eq:displacement_BCH}
\end{equation}
Its matrix elements in the Fock basis $ \{ |n\rangle \} $ read
\begin{align}
&\langle n | \hat{D}(\alpha) | m \rangle = e^{-|\alpha|^2/2}\nonumber\\ 
&\times\begin{cases}
e^{-|\alpha|^2/2} \sqrt{\dfrac{n!}{m!}} \, \alpha^{\,m-n} \, L_n^{(m-n)}(|\alpha|^2), & n \le m,\\[6pt]
e^{-|\alpha|^2/2} \sqrt{\dfrac{m!}{n!}} \, (-\alpha^*)^{\,n-m} \, L_m^{(n-m)}(|\alpha|^2), & n > m,
\end{cases}
\label{eq:Dnm_piecewise}
\end{align}
with the associated Laguerre polynomials $L^{(a)}_{n}(x)$, see, e.g., \cite{CahillGlauber1969}.
The complex parameter $\alpha$ encodes the position and momentum displacements. 
A phase-space shift by $x_0$ in the position and $p_0$ in the momentum coordinate corresponds to
\begin{align}
\alpha &= \frac{1}{\sqrt{2}}\!\left(\frac{x_0}{x_{\rm ho}} + i\,\frac{p_0}{p_{\rm ho}}\right),\nonumber\\
x_{\rm ho}&=\sqrt{\frac{\hbar}{m\omega}},\;\; p_{\rm ho}=\sqrt{\hbar m\omega}.
\label{eq:alpha_x0p0}
\end{align}
With these definitions a momentum-displacement reads $\hat D(i k) \;=\; e^{i \sqrt{2}\,k\,\hat x}$.
Therefore,
\begin{equation}
\cos\!\big(\sqrt{2}\,k_{\rm lat}\,\hat x\big)
=\tfrac12\!\left[\hat D(i k_{\rm lat})+\hat D(-i k_{\rm lat})\right].
\end{equation}
This allows us to express the harmonic-oscillator lattice matrix element as
\begin{equation}
S_{n,m}^{\rm HO}
=\frac{1}{2}\Big(\langle n|\hat{D}(-i k_{\rm lat})|m\rangle+\langle n|\hat{D}(i k_{\rm lat})|m\rangle\Big),
\label{eqn:lattice_HO_displacement_operator}
\end{equation}
which evaluates to Eq.~\eqref{HO_lattice_matrix_elements_main} after the substitution of \eqref{eq:Dnm_piecewise}. 

\section{Orthogonal transformation among the lab and center-of-mass frames}
\label{sec_app:Rotation_to_Lab_frame}
Considering that the relative and center-of-mass coordinates are connected by a rotation with the lab-frame coordinates, we can introduce the following relation
\begin{align}
\left. \begin{array}{rl}
     r&=\frac{1}{\sqrt{2}}(x_{\uparrow} - x_{\downarrow})  \\
     R&=\frac{1}{\sqrt{2}}(x_{\uparrow} + x_{\downarrow})
\end{array} \right\} \Rightarrow
\left. \begin{array}{rl}
     x_{\uparrow}&=\frac{1}{\sqrt{2}}(R + r)  \\
     x_{\downarrow}&=\frac{1}{\sqrt{2}}(R - r)
\end{array} \right\}\nonumber\\ \Rightarrow
\left(
\begin{array}{c}
   x_{\downarrow}  \\
   x_{\uparrow}
\end{array}
\right) =
\underbrace{\left(
\begin{array}{c c}
  \cos \frac{\pi}{4} & -\sin \frac{\pi}{4}  \\
  \sin \frac{\pi}{4} & \cos \frac{\pi}{4}
\end{array}
\right)}_{\equiv R_z(\frac{\pi}{4})}
\left(
\begin{array}{c}
   R  \\
   r
\end{array}
\right).
\end{align}
We know that the generator of rotation is the angular momentum, i.e. $\hat{R}_z(\theta) = e^{-\frac{i}{\hbar} \hat{L}_z \theta}$, with $\hat{L}_z  = \hat{x}_{\uparrow} \hat{p}_{\downarrow} - \hat{x}_{\downarrow} \hat{p}_{\uparrow} = i \hbar ( \hat{a}_{\downarrow}^{\dagger} \hat{a}_{\uparrow} - \hat{a}^{\dagger}_{\uparrow} \hat{a}_{\downarrow} )$. The key idea to proceed is to identify that the $\hat{S}_y$ operator in the Schwinger oscillator model of angular momentum that reads $\hat{S}_y= \frac{i}{2}(\hat{\alpha}_-^{\dagger}\hat{\alpha}_+ - \hat{\alpha}_+^{\dagger}\hat{\alpha}_-)$ \cite{Schwinger1952, Sakurai}. The requirements for the $\hat{\alpha}_{\pm}$, $\hat{\alpha}_{\pm}^{\dagger}$ operators of the Schwinger model are that they satisfy the standard commutation relations, $[\hat{\alpha}_{\sigma}, \hat{\alpha}_{\sigma'}^{\dagger}] = \delta_{\sigma, \sigma'}$, which are obviously satisfied by the operators $\hat{a}_{1,2}$, $\hat{a}_{1,2}^{\dagger}$ since they correspond to actual harmonic oscillator degrees-of-freedom. Therefore $\hat{L}_z = 2 \hbar \hat{S}_y$ by the identification $\hat{\alpha}_+ = \hat{a}_{\uparrow}$ and $\hat{\alpha}_- = \hat{a}_{\downarrow}$.

This results in a dramatic simplification of the physical situation, since it allows us to use the technique of the SU(2) spin-algebra to transform between the lab and center-of-mass frame. In particular, by the definition of the small Wigner d-matrix, $d^J_{M', M}(\theta)$ it is known that
\begin{equation}
    \langle J', M' | e^{-i \hat{S}_y \theta} | J, M \rangle = \delta_{J',J} d^J_{M', M}(\theta),
\end{equation}
where $J$ is the principal angular momentum quantum number and $M = -J, -J+1, \dots, J$ is its $z$ projection.
Therefore, in order to calculate the matrix elements of the transformation we are interested in, it suffices to identify how the spin states $|J, M\rangle$ are expressed in terms of the harmonic oscillator basis. Given the connection to the Schwinger model this identification is $| J , M \rangle = |\frac{n_{\uparrow} + n_{\downarrow}}{2},\frac{n_{\uparrow} - n_{\downarrow}}{2}\rangle$ and thus
\begin{align}
  \hat{U}^{\text{CM$\to$lab}}_{m,m';n,n'}&=\langle \tilde{\Psi}_{m,m'} | e^{- \frac{i}{\hbar} \hat{L}_z \pi/4} |\tilde{\Psi}_{n, n'} \rangle\nonumber\\
  &= \delta_{\frac{m+m'}{2}, \frac{n+n'}{2}} d^{\frac{n+n'}{2}}_{\frac{m-m'}{2}, \frac{n-n'}{2}}(\pi/2).
\end{align}
The connection to an effective spin also allows to calculate the $d^J_{M, M'}(\pi/2)$ efficiently since it can be seen as the unitary transformation that maps the eigenstates of $S_x$ in the $S_z$ basis and thus for any $J$ it can be easily calculated by diagonalizing the $(2 J +1) \times (2J+1)$ tridiagonal matrix $\hat{S}_x = \frac{1}{2} (\hat{S}_+ +\hat{S}_{-})$ and applying the Condon-Shortley phase convention to the resulting eigenvectors.

Therefore, it is convenient to work directly with the Wigner 
$d$–matrix elements. The transformation from CM coefficients 
$C^{\text{CM}}_{n,n'}$ to lab-frame coefficients $\kappa^{\rm lab}_{m,m'}$ then reads
\begin{equation}
\kappa_{m,m'}^{\rm lab}=\sum_{n,n'} \hat{U}^{\text{CM$\to$lab}}_{m,m';n,n'}
C^{\text{CM}}_{n,n'}.
\label{eq:kappa}
\end{equation}
This directly leads to the wavefunction in lab-frame coordinates \eqref{eqn:one_body_HO}.
\bibliography{bibliography}

\end{document}